\documentclass[10pt,journal,compsoc]{IEEEtran}

\usepackage{fancyhdr}
\usepackage[normalem]{ulem}
\usepackage[hyphens]{url}
\usepackage{graphicx}
\usepackage{subcaption}
\usepackage{listings}
\usepackage{lipsum}
\usepackage{courier}
\usepackage{array}
\usepackage[usenames, dvipsnames]{color}
\usepackage[sort,nocompress]{cite}
\usepackage[final]{microtype}
\usepackage{flushend}
\usepackage{xspace}
\usepackage{multirow}
\usepackage{tikz}
\usepackage{enumitem}
\usepackage{subcaption}

\usepackage[draft,bookmarks=true,breaklinks=true,letterpaper=true,colorlinks,linkcolor=black,citecolor=blue,urlcolor=black]{hyperref}
\usepackage{cleveref}

\title {Pre-cache: A Microarchitectural Solution to prevent Meltdown and Spectre}

\newcommand{\PerfImprvAvg}{2.3}
\newcommand{\PerfImprvMax}{22.7}
\newcommand{\PerfImprvMin}{-4.5}
\newcommand{\pc}{Pre-cache\xspace}
\newcommand{\md}{Meltdown\xspace}
\newcommand{\spt}{Spectre\xspace}
\newcommand{\KL}[1]{\textcolor{magenta}{KL: #1}}
\newcommand{\JMS}[1]{\textcolor{magenta}{JMS: #1}}
\newcommand{\HC}[1]{\textcolor{ForestGreen}{HC: #1}}

\newcommand{\ignore}[1]{}
\renewcommand{\HC}[1]{#1}

\usepackage{xstring}
\newcommand{\PP}[1]{
\vspace{2px}
\noindent{\bf \IfEndWith{#1}{.}{#1}{#1.}}
}

\lstdefinestyle{customc}{
  belowcaptionskip=0.2\baselineskip,
  breaklines=true,
  xleftmargin=\parindent,
  language=C,
  showstringspaces=false,
  basicstyle=\footnotesize\ttfamily,
  keywordstyle=\bfseries\color{green!40!black},
  commentstyle=\itshape\color{purple!40!black},
  identifierstyle=\color{blue},
  stringstyle=\color{orange},
  numbers=left,
}

\lstdefinestyle{customasm}{
  belowcaptionskip=1\baselineskip,
  frame=L,
  xleftmargin=\parindent,
  language=[x86masm]Assembler,
  basicstyle=\footnotesize\ttfamily,
  commentstyle=\itshape\color{purple!40!black},
}

\begin{document}

\author {
	\IEEEauthorblockN {
		Subhash Sethumurugan,
		Hari Cherupalli,
		Kangjie Lu,
		John Sartori \\
		\IEEEauthorblockA {
			University of Minnesota
			}
		}
	}

\IEEEtitleabstractindextext{
\begin{abstract}
Recent work has shown that out-of-order and speculative execution
mechanisms used to increase performance in the majority of today's
processors expose the processors to critical attacks.
These attacks, called Meltdown and
Spectre, exploit the side effects of performance-enhancing features
in modern microprocessors to expose secret data through side
channels in the microarchitecture. The well known implementations of these attacks exploit cache-based side channels since they are the least noisy channels to exfiltrate data. 
%
While some  software patches attempted to mitigate these
attacks, they
are ad-hoc and only try to fix the side-effects of the vulnerabilites. They may also impose a  performance overhead of up to 30\%.  In this paper, we present a microarchitecture-based solution for
Meltdown and Spectre that addresses the vulnerabilities exploited by the attacks. 
Our solution prevents flushed instructions from exposing data to the cache. Our approach can also be extended to other memory structures in the microarchitecture thereby preventing variants of the attacks which exploit these memory structures.
We further  identify two new variant attacks based on exploiting the side effects of speculative and  out-of-order execution and show how our solution can be used to
prevent these attacks. 
Evaluation results show that our microarchitectural solution not only restores secure out-of-order and speculative execution, 
but also has relatively low overhead and does not significantly impact performance for most applications.

\end{abstract}


}

\maketitle

\vspace{-10pt}
\section{Introduction}
\label{sec:intro}
\vspace{-1pt}
\IEEEPARstart{R}{ecently}, two critical security attacks have been reported
that affect a large fraction of microprocessors used
today~\cite{meltdown,spectre}. These attacks, 
\md and \spt, exploit fundamental vulnerabilities caused by
performance-enhancing mechanisms employed in modern processors.
\md exploits delayed privilege checking in processors that use out-of-order 
execution (OoOE), while \spt  exploits 
branch prediction and branch target prediction mechanisms. Both 
mechanisms allow speculatively-executed instructions to illegally 
access data from unauthorized memory locations. 
In \md~\cite{meltdown}, an instruction that is
executed out-of-order accesses data from the kernel memory
space before being invalidated. The attack then induces state
changes (i.e., side effects) in cache based on the value of the 
illegally-read data. While this instruction and subsequent
instructions are eventually squashed, the side effects remain in the
cache, which can be exploited to infer the illegally-read data using 
cache-based timing attacks such as
Flush+Reload~\cite{flush+reload}.
In \spt~\cite{spectre}, an instruction
that is speculatively-executed based on a branch prediction and/or a
branch target prediction reads data from unauthorized memory
locations before being squashed. Again, the attack induces side
effects in cache based on illegally-read data.
Even though the instruction is not allowed to commit, the side 
effects remain even after the
instruction is squashed, and thus can be accessed by a timing-based
attack on the cache.
While \md reads secret data in the kernel space, \spt can read 
sensitive data in the address space of another process.
In addition to \md and \spt, researchers have proposed several
variant attacks (\Cref{sec:attacks_and_variants}).

Both \md and \spt are practical and critical. 
They can efficiently leak secrets stored in the memory of
kernel space or other running programs, such as passwords,
personal photos, emails, instant messages, and business-critical 
documents. Researchers have shown how to dump memory, spy
passwords, and reconstruct
images~\cite{meltdown}.
\md and \spt impact a variety of computing devices,
including personal computers, mobile devices, and cloud computing nodes. Depending 
on the cloud provider's infrastructure, it may even be possible to steal data from other customers. 
Security researchers consider these vulnerabilities catastrophic
because they are fundamental and widespread. These vulnerabilities 
exist in fundamental mechanisms employed in most processors
manufactured in the last 20 years~\cite{smith98}. As such, they pose critical 
security threats to most computer devices, and fixing these 
vulnerabilities is urgent.
%
While reverting back to in-order execution is an
intuitive and effective solution, it would lead to serious 
performance degradation~\cite{Hennessy}. Similarly, disabling speculative
execution forfeits its significant performance benefits and is thus 
unacceptable~\cite{Hennessy}. Rather, an effective solution is one that allows 
out-of-order and speculative execution yet resolves the
underlying security problems introduced by these mechanisms.

While the vulnerabilities exploited by the attacks are in hardware, the attacks
themselves are implemented in software. Therefore, solutions can be based
on software and/or hardware.
Since the threats were made public, most processor manufacturers and major system software vendors have proposed solutions to stop these attacks 
(see \Cref{s:relwk}).
In general, software-based solutions address symptoms (e.g., the way in which
existing attacks exploit the hardware vulnerabilities) and thus are
ad hoc. For example, KAISER~\cite{kaiser} isolates the kernel
page table from user space, which stops \md  from accessing
kernel addresses directly from user space. This fixes
\md but not \spt or other variant
attacks~\cite{SgxPectre}. Moreover, it imposes 5\% performance overhead 
for most workloads and 30\% or more performance overhead in
some cases~\cite{corbet}.

Although hardware-based fixes have the potential to eliminate the vulnerabilities exploited by these attacks,
existing solutions fix only particular variants.
For example, hard-split bit (preventing access to kernel addresses from user space) suggested by the original \md 
paper~\cite{meltdown}, stops only \md.
Similarly, Intel recently developed a hardware
solution -- introducing new CPU instructions that eliminate branch
speculation -- to prevent the \spt variant (Spectre V2) that targets branch target injection.
Such a solution still leaves the processor vulnerable to the \spt variant that targets bounds check bypass (Spectre V1). 
We conclude that to prevent \md, \spt, and their current and future variants, we need to address the vulnerabilities in hardware; we require an efficient and general hardware-based solution that eliminates the vulnerabilities exploited by the attacks. 

%

In this paper, we propose a microarchitecture-based solution -- \pc -- to comprehensively 
prevent \md, \spt, and their variants that exploit the memory structures in the processor.
\pc prevents microarchitectural side effects caused by
out-of-order and speculative execution, and therefore, is able to
prevent the aforementioned attacks. This paper also emphasizes a 
security rule that designers should not overlook when evaluating 
future performance enhancements using memory structures for microprocessors. 
Namely, speculative execution should not leave side effects in microarchitectural state.
\pc prevents side effects of OoOE and speculative execution from being cached until they have been properly vetted. \pc is a buffer that 
holds a block fetched from memory until at least one instruction
requesting the block is committed. Meanwhile, the
\pc maintains spatial and temporal locality by making data
available for subsequent loads to the same block. Once at least
one instruction corresponding to the block in the \pc is
committed, the block is moved from \pc to cache. If no instruction
corresponding to the block is committed (i.e., squashed), the block
in the \pc is discarded, preventing it from being accessible
through a timing-based attack against the cache. Since all instructions of
a thread are either committed or discarded on a context switch, the associated
block in the pre-cache is either stored to cache or discarded,
preventing any timing-based attacks on the \pc. 
Since the pre-cache prevents data loaded by squashed instructions from entering the cache, it mitigates cache pollution, which may improve performance for some applications, as shown in \Cref{sec:results}.



This paper makes the following contributions.

\noindent $\bullet$ We present \pc as a microarchitectural solution that addresses the vulnerabilities exploited by \md and \spt. As opposed to existing software solutions, which
target specific variants,
\pc is a general solution that prevents variants of Meltdown- and Spectre-based attacks that use known memory structures,
by isolating the side effects caused by OoOE and speculative 
execution.
\pc can also be naturally extended to protect new memory structures
to stop future variant attacks.


\noindent $\bullet$ We present a general \pc microarchitecture that can be applied in multi-level cache hierarchies and multi-core systems. We describe how our implementation maintains inclusivity in multi-level caches and ensures coherence and consistency in multi-core systems. 

\noindent $\bullet$ We identify two new variants of \md and \spt that exploit instruction cache and show that \pc can prevent them. We further propose mechanisms to reduce traffic between the processor and \pc when protecting instruction cache.

\noindent $\bullet$  We show that in addition to addressing the vulnerabilities exploited by these attacks, \pc does not significantly degrade performance, and may even improve performance for some applications. We observed performance improvements of up to \PerfImprvMax\%, \PerfImprvAvg\% on average.


\vspace{-12pt}
\section{Related Work and Background}
\label{s:relwk}
\vspace{-1pt}

In this section, we first present the background required to explain \pc. This includes 
explaining out-of-order and speculative execution mechanisms, cache-based side channels, and the \md and \spt attacks and their variants. We then present recent patches against the attacks.

\vspace{-14pt}
\subsection{Out-of-order and speculative execution}
\vspace{-1pt}

One of the causes of the \md attack stems from the implementation of out-of-order
execution (OoOE) in processors. OoOE improves performance by using available instruction cycles to execute later instructions while
waiting for a preceding instruction to be completed.
As long as all required resources (e.g., operands) an instruction needs
are available, it can execute early, in parallel with the
execution of preceding instructions. 
The performance benefit of OoOE can be significant, since long-latency instructions would otherwise stall the processor, preventing forward progress.
OoOE strictly follows architectural specifications. That is, if the
execution of an instruction is deemed invalid (e.g., raising an
exception), the \emph{architectural} effects of the out-of-order 
instructions are discarded. However, the \emph{microarchitectural} side
effects can linger in the processor which are exploited by the \md
attack. For more details of OoOE, readers can refer to works on the \md
attack~\cite{meltdown} and the Tomasulo algorithm~\cite{Tomasulo}. 

The \spt attack exploits the speculative
execution mechanism which is another performance enhancment feature in most modern microprocessors. When execution reaches a conditional branch instruction whose target has yet not been resolved, speculative execution predicts the target and continues execution. If the prediction is wrong, the processor will abandon the execution of the wrong path and resume  execution of the correct one incurring a performance overhead. However, if the prediction accuracy is high, speculative execution can significantly improve a processor's performance. More details about speculative execution and branch prediction can be found in the \spt paper~\cite{spectre}.

\vspace{-6pt}
\subsection{Cache-based Side Channel Attacks}
\vspace{-1pt}
Although OoOE and speculative execution strictly follow 
architectural definition, they leave microarchitectural side effects in the processor that can open side channels that attackers can leverage to infer sensitive data. Some of the least noisy channels that have been studied extensively in literature are cache side channels~\cite{206170, gotzfried2017cache, 203183, SchwarzWGMM17, Liu:2015:LCS:2867539.2867673, flush+reload, cache-attacks, cache-missing}. In general, 
sensitive data is used to change the state of the cache. A change in cache state could be, for example, loading new cache lines or evicting loaded cache lines. Such a state change can be used by the attacker to infer the sensitive data. One way to do this is by measuring the timing differences when accessing different addresses that may be in the cache, thereby realizing which lines were loaded or evicted. This information can be used to infer the sensitive data. 
%

\PP{Preventing cache side channel attacks}
%
Caches are one of the least noisy side channels for exfiltrating data in \md and \spt. Existing defense mechanisms against
cache-side channel attacks can be classified into three categories. (1)
Cache partitioning~\cite{NoMo, CATalyst, pcache, STEALTHMEM}
prevents an attacker from probing the cache used by the victim. (2) 
Randomization~\cite{ZhenghongWang, StopWatch, Liu:2014}
introduces randomization to the attacker's inference of sensitive
information. (3) Detection~\cite{cloudradar, HexPADS} detects abnormal 
patterns, such as a significant cache miss rate caused by side channel
attacks.

In general, preventing side channel attacks is difficult. In fact, it is
impossible to exhaustively identify all potential side channels. 
Thus, existing
defense and detection mechanisms are for specific, known side channels.
Further, randomization and detection mechanisms are 
probabilistic and cannot guarantee security~\cite{cloudradar, ZhenghongWang, StopWatch, Liu:2014}, and 
cache partitioning typically incurs a significant performance
overhead~\cite{STEALTHMEM}. 
As such,  in this paper, instead of preventing cache side-channel attacks 
themselves, we aim to prevent side effects of OoOE and speculative 
execution from being cached.

\vspace{-6pt}
\subsection{\md, \spt, and their Variants}
\label{sec:attacks_and_variants}
\vspace{-1pt}
Since the discovery of the vulnerabilities caused by OoOE and
speculative execution, various attacks have been proposed. The \md
attack~\cite{meltdown} exploits OoOE to read privileged
data in kernel space and translate the read data into
microarchitectural side effects in cache that can be leaked
through cache side channels.  The \spt attack~\cite{spectre} exploits speculative
execution to leak data in the user space of other processes.
Researchers have also demonstrated 
a few variants of the \spt attack: reading kernel memory from user space and reading host memory
from KVM guests~\cite{horn}, Prime+Probe variant of 
Spectre~\cite{meltdownprime}, and a Spectre attack targeting Intel
SGX~\cite{SgxPectre}. Other variants discovered by researchers include Rogue System Register Read~\cite{RSRE}, Speculative Store Bypass~\cite{SSB},  Lazy FP State Restore~\cite{LFSR}, and Return Stack Buffer (RSB). 
%
In fact, any new means of translating sensitive data into side effects or any new side channel for observing the side effects can lead to a new variant attack. 
In this paper, we address the most widely used side-channel for existing attack variants -- cache. We further show that our solution can be be extended to any memory structure in the micro-architecture.

\vspace{-6pt}
\subsection{Mitigation of \md and \spt}
\label{s:patches}
\vspace{-1pt}

KAISER~\cite{kaiser} mitigates the \md attack by changing the
operating system kernel code to isolate kernel memory from user space accesses. 
KAISER does not fix the lingering microarchitectural side effects of out-of-order execution -- but stops the way the \md
attack exploits out-of-order execution to leak information. 
The runtime performance overhead of KAISER can be significant -- about 5\% for most workloads and 
up to 30\% in some cases, even with the PCID (process-context 
identifiers) optimization~\cite{corbet}.
The original \md paper~\cite{meltdown} also suggests three hardware
solutions. (1) \emph{Completely disabling OoOE}: This will
result in unacceptable performance degradation as well as poor energy efficiency in
modern CPUs and thus is not viable. (2) \emph{Serializing the permission check for memory-read instructions}: This
will also introduce a significant performance overhead\cite{meltdown}.
(3) \emph{Introducing a new hard-split bit in a CPU control register}:  If
the hard-split bit is set, the kernel is relegated to the upper half
of the address space, and the user space must reside in
the lower half of the address space. Similar to KPTI, such a solution
is effective but only stops the \md attack. It does not eliminate all the vulnerabilities and also
cannot stop the \spt attack.

Researchers have also proposed some other partial solutions to prevent attack variants, such as  Google's ``Retpoline'', which handles speculative execution on indirect jumps, and Mozzila's reduction of resolution of JavaScript timers.
Such patches are reactive solutions for some known ways in which attackers have exploited
speculative execution. As such, attackers can propose 
variant attacks to bypass these patches.

SafeSpec~\cite{safespec} uses a shadow structure to prevent speculatively accessed data from reaching the cache before the speculation is resolved. The memory sub-system populates the shadow structure with cache line accessed by a speculatively executed instruction. After the speculation is resolved, the cache line is removed from the shadow structure and sent to the cache. If the speculation was invalid, the cache line is removed from structure but is not sent to cache. However, SafeSpec considers only single-core processor; extending SafeSpec to multi-core processors is difficult because of the following reasons. Coherency updates to cache lines in the shadow structure will be lost because the directory at the last level cache is unaware that the core that executes the speculative instruction has loaded the corresponding cache line in its shadow structure. Also, after the speculative instruction commits and the corresponding cache line is moved from the shadow structure to the cache, coherency updates must be sent to the other cores. In such cases, handling the race conditions between coherency updates from other cores and the shadow structure is non-trivial.
Noted that, the lower-level cache update from the shadow structure could affect the inclusive nature of the cache hierarchy; this is not accounted for in the Safespec design. In our \pc design, we maintain the inclusive nature of the cache hierarchy and provide support for multi-core architectures through detailed design of \pc directories at every level of cache. 

Invisispec~\cite{invisispec} uses a speculative buffer (SB) to hide the cache effects of a speculative load until the load is determined to be safe. The speculative buffer contains the cache lines accessed during a speculative load. Since the speculative buffer is invisible to the memory sub-system, invalidations to the cache lines in the speculative buffer do not reach the core. To avoid consistency issues, the design issues two memory accesses for the same load -- one to populate a SB entry and the other to validate or expose the effects of the load. In our pre-cache design, we avoid consistency issues by allowing the invalidations to reach the pre-cache by maintaining pre-cache directories at every level. The pre-cache directories keep track of pre-cache entries without affecting the cache or directory states. This allows the memory sub-system to maintain consistency without an additional load and without suffering significant performance degradation. 

Note also that InvisiSpec requires significantly more design changes than Pre-Cache. Implementing InvisiSpec requires the following modifications.
(1) InvisiSpec requires additional logic to determine if the instructions ahead of the current instruction in the ROB are non-squashable, status bits added in load queue such as State, Performed, Prefetch, etc. These modifications could demand higher verification overhead, since they involve multiple modules in the execution of a secure load. 
(2) In InvisiSpec, a load can have multiple manifestations. That is, a load may or may not be put in the SB based on whether it is considered safe by address resolution. Similarly, a load in the SB can be made visible after it is considered safe, either by validation or exposure. These variants introduce additional verification overhead and increase design time.

Compared to InvisiSpec, our solution is simpler and more modular.
All loads are assumed to be unsafe, and their data are put in \pc. We only monitor the commit signal and use the load address to determine if data can be made visible to the cache hierarchy. We maintain consistency and coherence using \pc directories. Since \pc directories are modeled after existing cache directories, verifying a \pc-based design is a straightforward extension of the original design verification, since testbenches and methodologies for verifying directories already exist.

\vspace{-6pt}
\section{Pre-cache}
\label{sec:design}
\vspace{-1pt}

As discussed in \Cref{sec:intro}, a major cause for \md and \spt is that
transient instructions (instructions that do not get committed) can leave side effects in the microarchitecture of a processor. In the most popular variants of the attacks, the side effects manifest as cache pollution.
This leads to a fundamental security vulnerability in
microprocessors, where values loaded into the cache by transient
instructions can be read through side-channel attacks.
In this section, we first abstract the attacks (the variants we target in this paper) to capture their essential steps and then present our solution -- \pc -- in detail.

\vspace{-14pt}
\subsection{Abstracting the Attacks}

\begin{figure*}
\begin{minipage}[c]{0.65\textwidth}
\includegraphics[width=\linewidth]{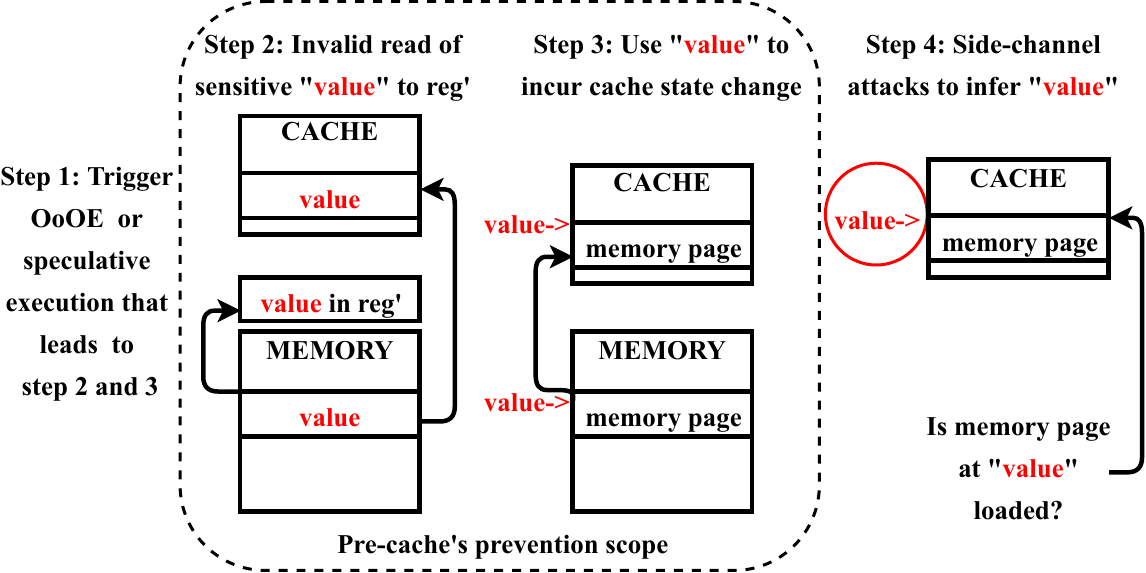}
\caption{Abstraction of \md and \spt attacks: Step 2 and step 3 execute transient instructions allowed by OoOE or speculative execution. \pc prevents side effects in both step 2 and step 3 by containing the data and side effects in an isolated buffer, making step 4 impossible.}
\label{fig:attack-abstract}
    \end{minipage}%
    \hspace{10pt}
    \begin{minipage}[c]{0.33\textwidth}
	\centering
\includegraphics[width=\linewidth]{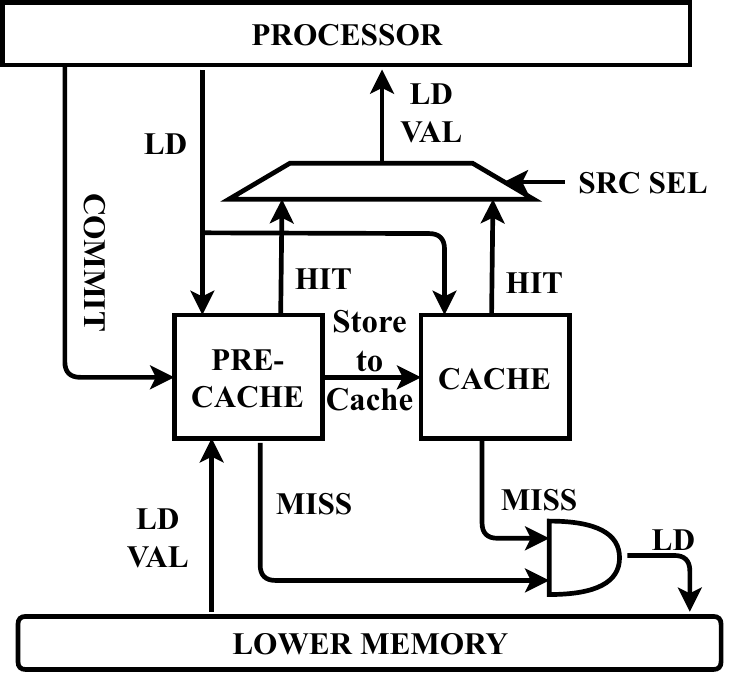}
\caption{\pc buffers speculatively-loaded data until speculation is resolved. If the loading instruction commits, data is sent to cache; otherwise, discarded.}
\label{fig:pre-cache-overview}
\end{minipage}
\vspace{-20pt}
\end{figure*} 

\vspace{-1pt}
Since \md and \spt have several variants (Speculative Store Buffer Bypass, Lazy FPU State Restore, SgxPectre, etc.) that rely on side-effects in memory structures in the processor's microarchitecture, we
first abstract general and essential steps of the attacks so that we can propose an effective solution to prevent all such variants. 
\Cref{fig:attack-abstract} shows four essential steps in these
attacks. Step 1 triggers either OoOE or speculative execution, which leads to step 2 and 3. Step 2 is an invalid memory
read that loads a secret value from memory to a register, which also
loads from the target memory page to cache if a cache miss occurs. Since attackers cannot directly read the secret value in the register or in
the cache, step 3 indirectly translates the secret value into cache
state changes (i.e., side effects), for example, by using it as a data/code pointer. In this way, the memory address 
pointed to by the secret value will be loaded to cache if it is absent. The last step leverages side channel attacks to infer the 
secret value by checking if the memory address pointed to by the secret value is loaded in cache. 

Stopping any step in \Cref{fig:attack-abstract} can prevent \md and
\spt (the cache based variants). However, two requirements for the solution are (1) good
performance and (2) completeness. Stopping step 1 (i.e., disabling
OoOE and speculative execution) is unacceptable, due to significant
performance overhead~\cite{Hennessy}. Stopping step 4 is difficult because it is not 
possible to eliminate all side channels. Thus, we choose
to stop steps 2 and 3.
One might think that a simple means of attacking the problem at its
source could be to invalidate the cache lines accessed by squashed instructions.  However, this would create a new problem; if the cache line is already in the cache, a squashed instruction might trigger an invalidation that was not solicited. An attacker could then exfiltrate data by tracking which addresses are slower to read instead of faster. All these considerations lead us to propose \pc: containing the side effects of OoOE and speculative
execution and preventing the side effects from entering the  cache if the execution is deemed invalid.


\vspace{-5pt}
\subsection{Threat Model}
\label{sec:threat_model}
\pc prevents the \md and \spt attacks that uses memory structures as side channels. As shown in
\Cref{fig:attack-abstract}, both attacks follow several essential
steps to finally leak secret values.
Stopping any step can prevent the attacks. We choose to prevent the attacks in the second and the third steps in \pc. 
That is, it contains side effects of transient instructions and
erases them upon invalidations of the instructions. 

%
%
We assume powerful attacks. We assume that attackers can run
arbitrary instructions in the victim machine, and victim software 
programs may also contain arbitrary instructions for accessing
secret data.
Also, attackers can freely control the branch predictor to
exploit speculative execution of any gadgets in the victim process. 
We assume the attackers are in user space, and their goal is to 
leak secret data in OS kernels or the user space of another process
through hardware side channels using side-channel attacks, such as
Flush+Reload or Prime+Probe.
While the attacks do not exploit OS-kernel vulnerabilities,
\pc's effectiveness does not rely on any specific feature or on the 
trustiness of OS kernels.
We assume that the processor has multiple cores that support 
simultaneous multithreading and an inclusive cache hierarchy.
While we assume that the attack code can be executed in a single-threaded, multi-threaded, or multi-core setting, the attack is limited to side channels in memory structures such as cache and TLB.


%
%
\pc focuses on preventing cache-based side-channel attacks, so other
channels such as port contention, bus contention, random number generators, AVX, DRAM row channel, power and electromagnetic signals are out of scope. Similarly, attacks such as spoiler~\cite{islam2019spoiler} that are based on speculative execution and try to reverse engineer virtual-to-physical mappings are out of scope.
We also assume that attackers do not have physical access to the
victim machine. 
While the approach of \pc is general, the \pc 
implementation discussed herein
contains side effects of transient instructions in major buffers, 
such as cache and TLB. We 
discuss how to extend \pc to protect other buffers without
significant challenges in \Cref{s:other-variants}.

\vspace{-6pt}

\vspace{-6pt}
\subsection{\pc Microarchitecture}
\label{sec:pc_microarchitecture}
\vspace{-1pt}

\ignore{moving this figure next to another to save space. if it doesn't work, put it back here.
\begin{figure}[ht]
\centering
\includegraphics[width=0.8\columnwidth]{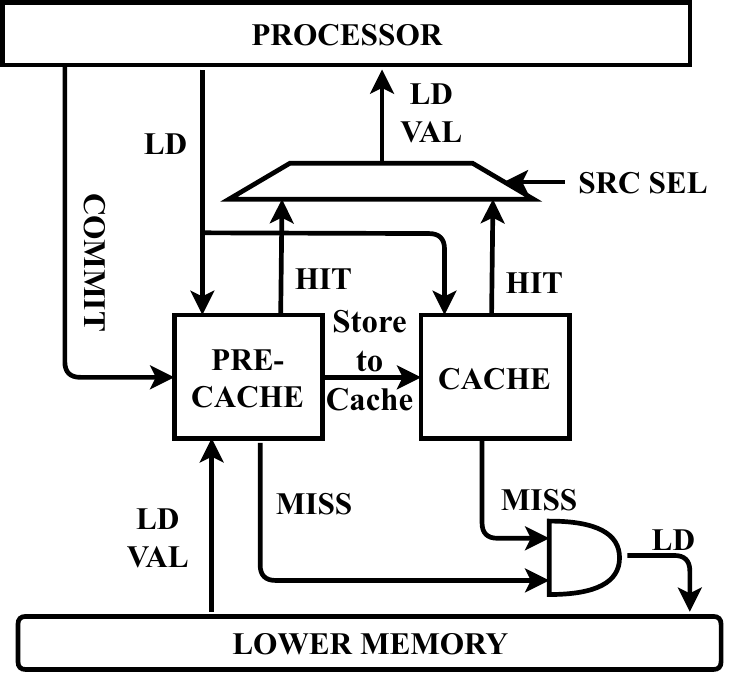}
\caption{\pc is a structure that buffers data retrieved from memory until the speculative nature of the data can be resolved. If the instruction that loaded the data commits, data is sent from \pc to cache. If the instruction is squashed, data in the \pc are discarded.}
\label{fig:pre-cache-overview}
\vspace{-10pt}
\end{figure}
}

\begin{figure*}[ht]
\centering
\begin{subfigure}[b]{0.26\textwidth}
\includegraphics[width=\linewidth]{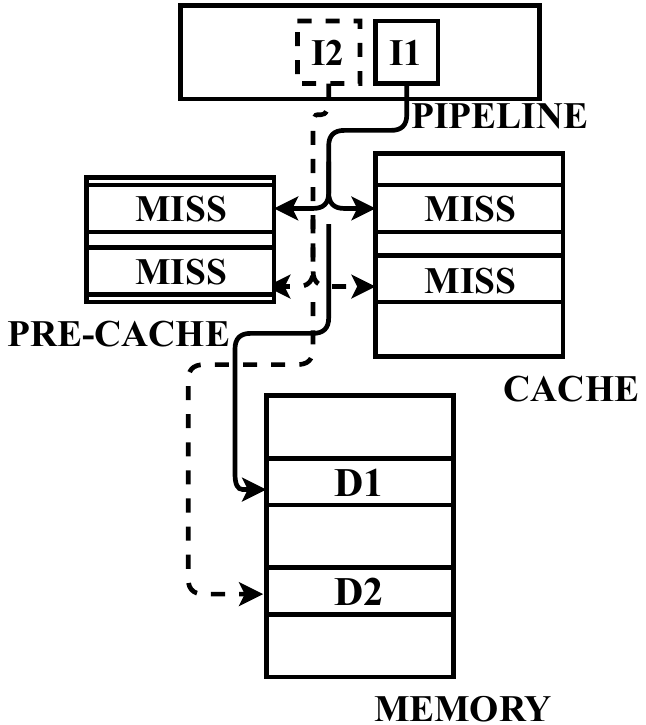}
\vspace{-14pt}
\caption{}
\label{fig:data-pre-cache-1}
\end{subfigure}
%
\begin{subfigure}[b]{0.212\textwidth}
\includegraphics[width=\linewidth]{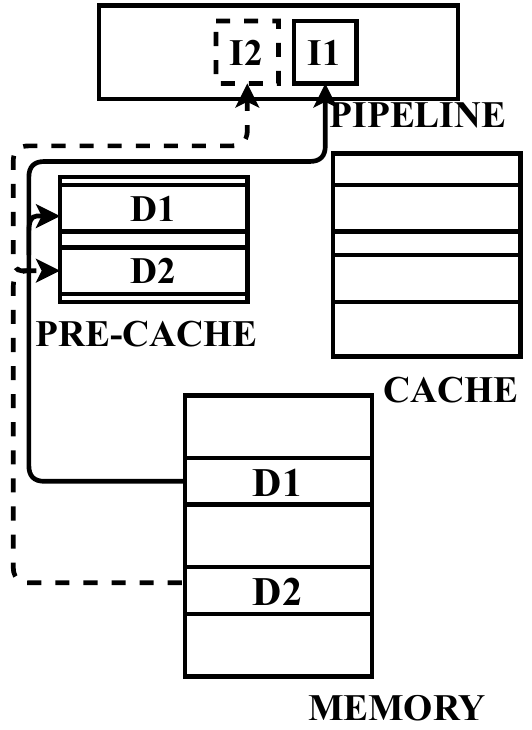}
\vspace{-14pt}
\caption{}
\label{fig:data-pre-cache-2}
\end{subfigure}
%
\begin{subfigure}[b]{0.266\textwidth}
\includegraphics[width=\linewidth]{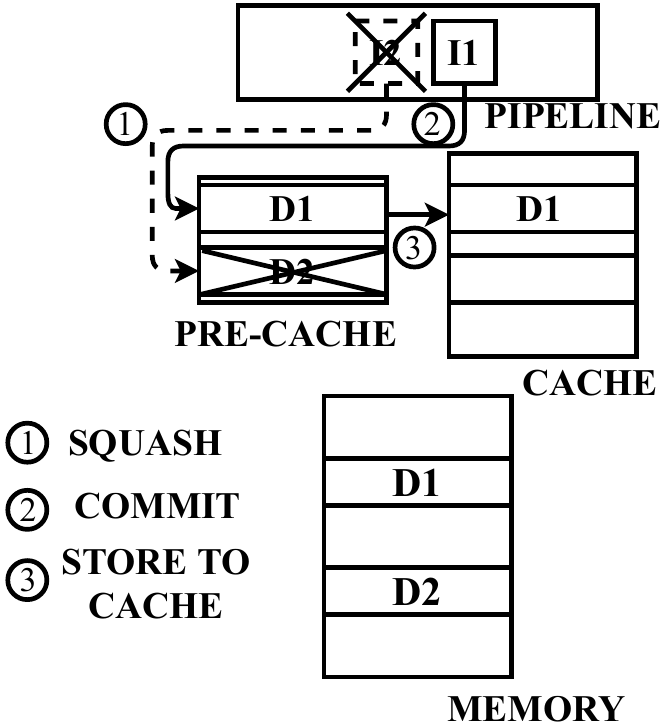}
\vspace{-14pt}
\caption{}
\label{fig:data-pre-cache-3}
\end{subfigure}
\includegraphics[width=0.216\linewidth]{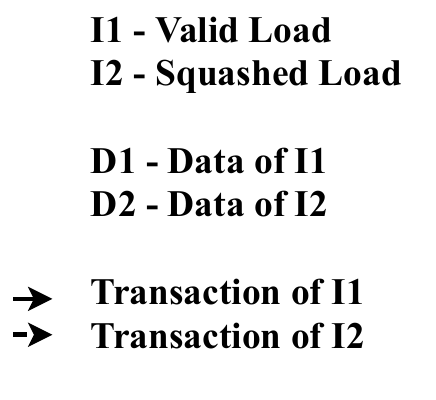}
\vspace{-10pt}
\caption{In the \pc architecture, data loaded from memory are buffered in the \pc until the instructions that loaded the data are confirmed to be valid. Data from transient instructions is not permitted to enter the cache.}
\label{fig:data-pre-cache}
\vspace{-10pt}
\end{figure*}

To fix the security vulnerabilities introduced by speculative and OoO execution that involve memory structures, we introduce a microarchitectural structure called \pc that prevents transient instructions from leaving side effects in the processor. Once these side effects are eliminated, \md and \spt attacks 
that use memory structures as a side-channel 
are averted, because the sensitive data they target is never exposed to the memory structures. 
\Cref{fig:pre-cache-overview} provides an overview of how \pc fits into the processor microarchitecture. 
We implement \pc as a buffer that sits alongside the L1 cache and temporarily holds data loaded from memory until it can be properly vetted, at which point it is either transferred to the cache or discarded. On a memory load, the address is supplied to the cache and \pc, which perform lookups in parallel. If either structure has a hit, the data is sent back to the processor. If both the cache and \pc miss, data is fetched from lower levels of memory (\Cref{fig:data-pre-cache-1}). Instead of placing retrieved data in cache, they are placed in \pc. Subsequent reads to the same cache line can be serviced by the \pc (\Cref{fig:data-pre-cache-2}).  When a load instruction is committed, its address is sent to the \pc, which performs a lookup; if \pc holds the cache line that the address maps to, 
that line is sent to cache. If any squash occurs, however, the contents of the \pc are cleared (\Cref{fig:data-pre-cache-3}). \Cref{fig:data-pre-cache} shows an example of \pc operation in which instruction I1 is a valid load that gets committed while instruction I2 is a transient load that gets squashed. Thus, the corresponding data D1 and D2 are handled differently. D1, which was loaded by committed instruction I1, is moved to the cache using an operation called \textit{store to cache (STC)}. Upon the squash of I2, however, the contents of the \pc are cleared, including D2. Barring data loaded by transient instructions from entering the cache and invalidating it in the \pc ensures that subsequent instructions cannot observe the data through a timing-based attack on the cache or \pc.

The \pc can be implemented as an associative array that stores the blocks that are fetched from  memory. By storing the blocks (like a cache), the \pc provides the same spatial locality as a cache. Similarly, allowing the blocks to be read from the \pc ensures temporal locality of the data.
The \pc only needs enough entries to buffer the data of in-flight loads, so it can be significantly smaller than the L1 cache. Evaluation with CACTI~\cite{cacti} reveals that the access latency for a 32-entry \pc (same size as our load queue) is roughly three times shorter than that of our L1 cache (32KB, 4-way); 
thus, even with a smaller cache or larger load queue, the access latency of the \pc can effectively be hidden when it is accessed in parallel with the L1 cache.

\vspace{-6pt}
\subsection{Multi-level Cache and Cache Coherence}
\label{sec:cache_coherence}
\vspace{-1pt}



For clarity, we have explained the design of \pc in the context of a single-level cache and single-core processor. 
In this section, we discuss how \pc design extends to a multi-core, multi-level cache hierarchy and how our design ensures cache coherence in a multi-core processor. 

\vspace{-6pt}
\subsubsection{Multi-level Cache}
\label{sec:multi_level_cache}
\vspace{-1pt}

In this section, we explain microarchitecture modifications to support \pc in a multi-level cache. For completeness, we assume an inclusive cache hierarchy; \pc for exclusive caching only requires a subset of the provisions for inclusive caching. 
In a processor with a multi-level cache hierarchy, the \pc is still placed in parallel with the L1 cache.
\pc serves the same fundamental role, 
but when a load is committed, the corresponding  block is sent from the \pc to all the cache levels that returned a miss when they were searched during the load. The STC traverses the levels of cache to obtain locks at all levels it will write to and performs writes after obtaining all necessary locks.
To achieve this functionality, we (1) maintain hit-level information in \pc, and (2) maintain \pc directories in the cache hierarchy.

\noindent {\bf Hit-level information in \pc}:
Since a STC only needs to write to the cache levels that missed during the load, the \pc keeps track of the level of memory where the load hit. When a load finds the block in a certain level, the ID of the level is recorded with the block in \pc. After a load commits, this ID is used by the STC to identify the levels of memory where the block must be written. For example, in a three-level hierarchy, we label L1, L2, L3, and Memory with the IDs 0, 1, 2, and 3. If a load hits in L2, an ID of `1' is buffered with the block in \pc. When the load commits, a STC is issued to write the data to L1.

\begin{figure}[ht]
\centering
\begin{subfigure}[b]{0.22\textwidth}
\includegraphics[width=\linewidth]{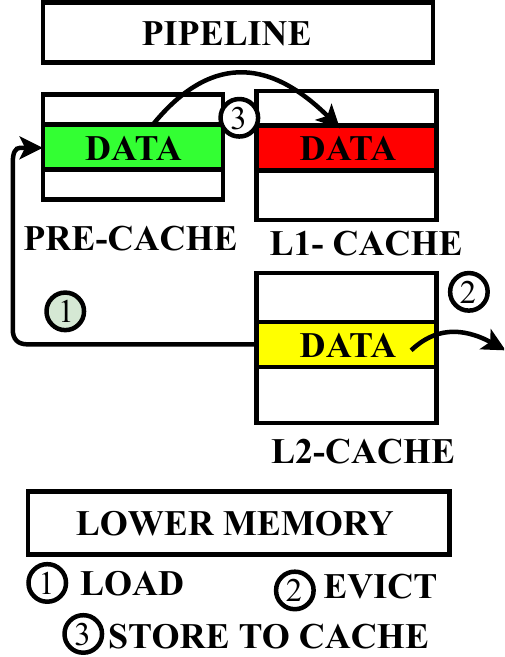}
\caption{No \pc directory}
\label{fig:multi-level_inclusivity_wrong}
\end{subfigure}
\begin{subfigure}[b]{0.224\textwidth}
\includegraphics[width=\linewidth]{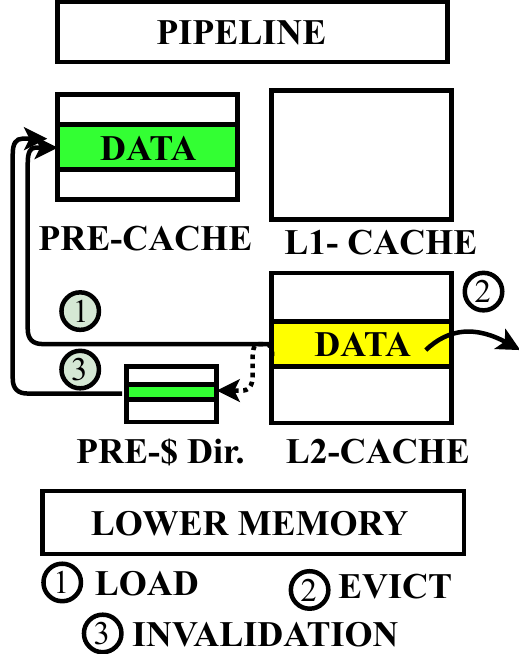}
\caption{With \pc directory}
\label{fig:multi-level_inclusivity_right}
\end{subfigure}
\vspace{-10pt}
\caption{\pc directories are necessary for inclusive cache. (a) 
If data in L2 is evicted before STC writes to L1, inclusivity is compromised. (b) 
The eviction at L2 causes an invalidation in \pc, maintaining inclusivity.}
\label{fig:multi-level_inclusivity}
\vspace{-12pt}
\end{figure}

\noindent {\bf \pc directories}
\pc directories are used to maintain inclusivity for a single-core multi-level cache and coherence in multi-core caching.
Consider a case where a block in \pc is evicted from the cache level where its load hit (due to a conflict miss) before the STC of the load populates the missed caches. For example, in the scenario described above, what if the block gets evicted from L2 before the STC reaches L1 (\Cref{fig:multi-level_inclusivity_wrong})? Inclusivity could be violated if the STC is allowed to complete. 
One way this inclusivity problem is solved in traditional cache systems is by maintaining directories to invalidate higher-level cache entries when a block is evicted at a lower level. 
Similarly, we supplement the cache directory at each level with a \pc directory that contains the addresses of the blocks that were loaded from that level and lower. Thus, \pc directories are \emph{reverse inclusive}, enabling invalidations from lower levels to be routed to the correct \pc in a higher level. In the example above (\Cref{fig:multi-level_inclusivity_right}), upon eviction at L2, the \pc directory at L2 is consulted, revealing that an invalidation should be sent to \pc to prevent the block buffered there from being written to L1.
Maintaining directories also affects the working of STC. Instead of traveling to the lowest cache level that a load missed from, a STC should travel to the cache level where the load hit (one extra level), to evict the \pc directory entry of the block in that level before proceeding to the higher-level caches to write data and evict \pc directory entries.


\vspace{-6pt}
\subsubsection{Multi-core \pc}
\label{sec:multi_core}
\vspace{-1pt}

\begin{figure}
\centering 
\includegraphics[width=0.85\linewidth]{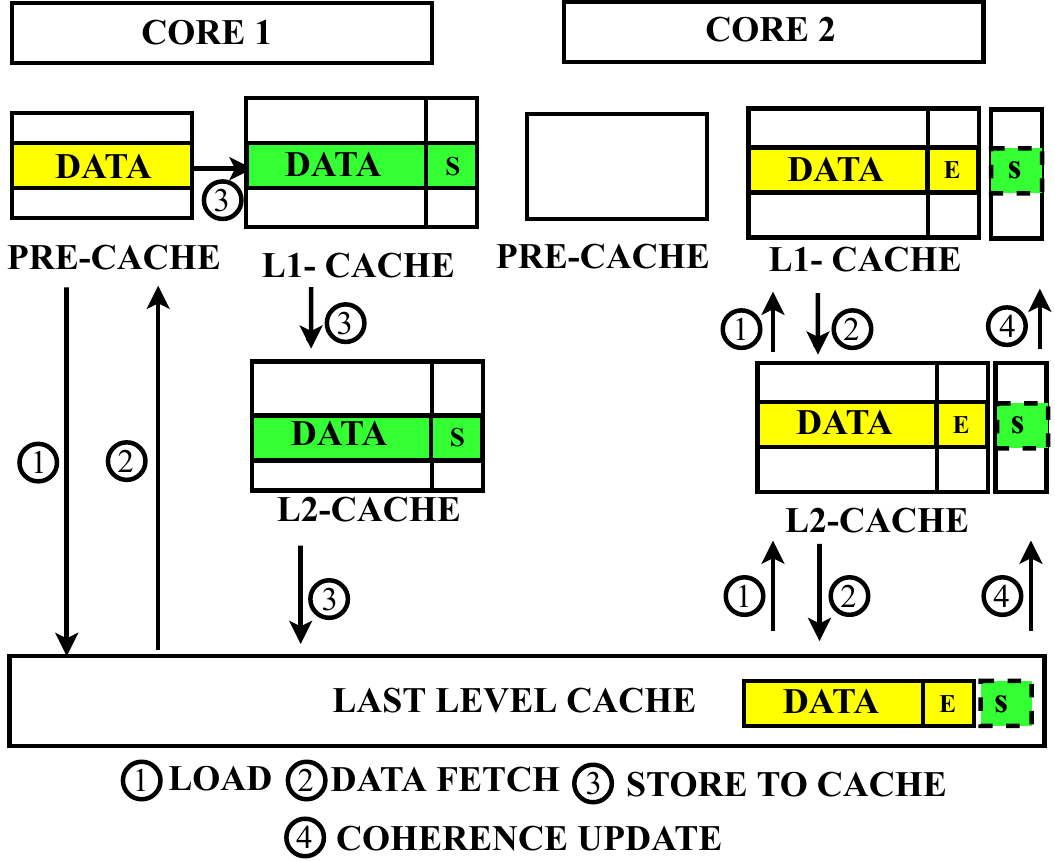}
\vspace{-10pt}
\caption{Coherence state in a \pc design is updated during STC.}
\label{fig:multi_core_coherence_update}
\vspace{-14pt}
\end{figure}

In a multi-core cache, \pc must maintain memory consistency and coherence.
\pc does not affect the ordering of committed loads and stores as seen by the system (see \Cref{sec:correctness}), and thus does not affect consistency. 
%
To ensure coherence in a design that supports \pc, we delay updating cache coherence states until after a load has committed, that is, during STC. This eliminates the possibility of using a speculative load to cause side effects based on coherence states. We explain this below with an example, illustrated in \Cref{fig:multi_core_coherence_update}.

Consider a load that finds its block, which is exclusive to another core, in L3. The load brings the block to the \pc  without updating the coherence state of the memory system. At commit, the corresponding STC locks entries in L1 and L2 of the same core before reaching L3. Consulting the directory of the L3 and finding that the block is exclusive to another core, the L3 would issue a state update request (from exclusive to shared) to the other core. The STC would wait for the state update to return to L3 before updating the L2 and L1 caches of its parent core, ensuring cache coherence while also preventing potential attacks based on cache coherence.

When a core issues a store to a block, all copies of the block in other cores must be invalidated, including Pre-cached copies. Invalidation of a block is handled in the same way as eviction (described in \Cref{sec:multi_level_cache}).
Invalidations must complete before the store can commit. This ensures that any subsequent load in another core does not read stale data, thereby preserving coherence. 
If a core is performing a STC for a block when an invalidation arrives, the STC is aborted during lock acquisition, 
upon reaching the cache level to which the invalidation has progressed. The abort causes any locks obtained by the STC to be released, and no data is written by the STC.

\ignore{
Consider a two-core processor with a three-level cache hierarchy and a shared L3 as LLC. Core 0 issues a store to an address that is in the L1 of Core 1. This causes the L3 to send an invalidation to Core 1's L2, which would send an invalidation to L1 if the L2 directory indicates that the block is in L1. Then, L2 would invalidate its own copy. In a \pc system, the \pc directory at L2 would also be consulted, and if resident, the block would be invalidated from \pc. For example, if a block was resident in the \pc of Core 1 and it had been loaded from L3, then the \pc directories in L2 and L3 would contain an entry for the block. Thus, an invalidation would be sent to the \pc after consulting the \pc directories of L3 and L2, in that order. The \pc entry must be invalidated before the store is allowed to commit. This ensures that any subsequent load in Core 1 does not read stale data, thereby preserving coherence. 
If the first core is performing a store-to-cache for a block when an invalidation arrives, the store-to-cache is aborted during lock acquisition, 
upon reaching the cache level to which the invalidation has progressed. The abort causes any locks obtained by the STC to be released, and no data is written by the STC.

}


\subsection{Other Cache Inclusion Policies:}

\HC{We did in inclusive because it is the most complex policy and highest overhead. We showed evaluations in inclusive policy because that is the most conservative and thorough in our explanations. We also show how to do it in other policies. 
So far the design of the Pre-cache design has be illustrated for an inclusive cache hierarchy. However, cache hierarchies can also be exclusive or non-inclusive non-exclusive (NINE).} 

\HC{In an exclusive cache hierarchy, data is placed in a cache as a result of evictions from the higher level. In a \pc augmented design only data corresponding to committed loads are placed in L1. Since L2 or lower levels receive data as a result of evictions in L1, data placed in L2 or lower levels are not transient. }

\HC{In a non-inclusive non-exclusive, data is brought in an inclusive manner, that is, data from memory is filled up at all levels of cache. However, an eviction from a cache doesn't trigger invalidations to its higher level caches. For a \pc augmented design, this further reduces the required hardware as \pc directories are no longer essential. }

\vspace{-6pt}
\subsection{Pre-caching Instructions}
\vspace{-1pt}

Processors read both data and 
instructions from memory. Like data, instructions are also stored in
the cache hierarchy. 
A timing-based side channel attack could also target an instruction cache to compromise security.
%
In this section, we present new variants of \md and 
\spt -- \textit{i\md}  and \textit{i\spt} -- that use the instruction cache to leak secret 
data from unauthorized memory locations. We then present a \pc design that prevents i\md and i\spt.

\vspace{-6pt}
\subsubsection{i\md and i\spt}
\label{sec:iattacks}
\vspace{-1pt}


Our i\md and i\spt attacks use instruction cache to leak secret data. 
%
\Cref{l:imd} illustrates i\md. Line 4 is an illegal
memory read that loads a secret value from kernel address space to register
\texttt{al}. Line 5 shifts \texttt{rax} to translate the
value to a page-level offset (multiple of 4096). Line 7 computes a code pointer (in
contrast to a data pointer in the original \md attack) and 
deferences it to jump to the resulting location in the code. This loads the code page
pointed to by \texttt{rbx+rax} to cache, making instructions in the page faster to access than others. After that, an attacker can launch side-channel attacks (e.g., Flush+Reload~\cite{flush+reload}) on the instruction cache to infer the value of the code
pointer and thus the secret value.

\Cref{l:ist} illustrates i\spt. Speculative execution can misspeculate the branch in line 1 and allow i\spt to bypass the bound check. Line 2
then performs an out-of-bounds read that may read sensitive data. Line
3 translates the read value into a page-level function pointer
(i.e., a code pointer). When the function pointer is dereferenced at
line 3, the code page is loaded into cache. By figuring out the
location of the code page using a side channel attack, an attacker can
infer the sensitive data. As in the i\md attack, our i\spt attack generates side effects
in the instruction cache rather than the data cache.

\begin{minipage}[t]{0.42\columnwidth}
\vspace{-18pt}
\begin{lstlisting} [caption=Core instruction sequence for i\md,
label=l:imd, captionpos=t]
;rcx = kernel address
;rbx = probe array
retry:
mov al, byte [rcx]    
shl rax, 0xc          
jz retry              
jmp qword [rbx + rax] 
\end{lstlisting}
\end{minipage}
\hspace{20pt}
\begin{minipage}[t]{0.4\columnwidth}
\vspace{-18pt}
\begin{lstlisting} [caption=Example gadget for i\spt,
label=l:ist, captionpos=t]
if (x < array_size)      
  foo = array[x] * 4096; 
  foo();                 
\end{lstlisting}
\end{minipage}

A critical step of the \spt attack is to find a proper gadget that,
similar to the one shown in \Cref{l:ist}, consists of a conditional
or indirect jump, a memory read that obtains a secret value, and a memory 
access based on the secret value. Our i\spt variant adds additional gadget configurations to the attacker's toolbox, significantly increasing the flexibility for attackers to find appropriate gadgets for \spt attacks.

\vspace{-6pt}
\subsubsection{Instruction \pc Microarchitecture}
\vspace{-1pt}

The high-level idea of pre-caching instructions is the same as
pre-caching data. The instruction \pc is implemented as a parallel access structure to the instruction cache. When an instruction commits, its address can be used to determine if any blocks should be propagated from \pc to cache.
However, it is not necessary to signal the i\pc for all committed instructions.
Fetching an instruction based on the value of secret data requires an indirect jump instruction. Since instructions in the basic block following an indirect jump share the speculative state of the jump, the instructions in a basic block can be committed to cache or discarded in bulk. When an indirect jump is detected in the decode stage, instructions are fetched from the basic block beginning at the speculated target given by the branch target buffer and brought into the i\pc. If a jump target was mispredicted, then the instruction \pc is cleared and the effects of speculative instruction fetch are not sent to the instruction cache. If a jump instruction commits, the instructions corresponding to the basic block following the jump are transfered from the i\pc to the cache, since they are no longer speculative. This approach reduces traffic between commit and i\pc, which can reduce power.

A challenge in implementing i\pc is handling multiple successive indirect jumps. When a jump commits, only the instructions up to the next indirect jump can be deemed non-speculative and moved to the instruction cache. To achieve this functionality, we can maintain a counter in decode stage to count the number of instruction cache blocks that have been fetched since the last indirect jump. When the next indirect jump is decoded, the counter value is enqueued in a circular buffer in the i\pc, telling how many cache blocks to transfer to the instruction cache when the next indirect jump is committed. An index to the first cache block containing the oldest indirect jump in the i\pc is also maintained. When an indirect jump is committed, the next counter value is dequeued and added to the current index value so that it now points to the block containing the next indirect jump in \pc.

\vspace{-6pt}
\subsection{\pc for Other Variants}
\label{s:other-variants}
\vspace{-1pt}

Our instruction-based variants of \md and \spt show that after
exploiting OoOE and speculative execution to illegally read secret data, an attacker 
has multiple possible locations in which to stash the secret data for later extraction.
In addition to data cache and instruction cache, many processors have
other performance-enhancing memory structures such as prefetchers, return stack buffer (RSB), and translation lookaside buffer (TLB). 
All these buffers have potential to cache side effects that can be
leaked through side channel attacks. 
\pc is a general mechanism that protects buffers from leaking
side effects introduced by transient instructions.
In the rest of this section, we discuss how to leverage
\pc to prevent variant attacks.

Side effects in TLB can be prevented using a TLB \pc. A TLB \pc buffers address translations for addresses that missed in TLB and resulted in a page table walk. An address translation in TLB \pc is sent to TLB only after the instruction that accessed the corresponding address commits. At the commit stage, the address accessed by each instruction is sent to the TLB \pc. If the address hits in the TLB \pc, its translation is sent to the TLB, ensuring that the TLB only contains address translations initiated by committed instructions. On a pipeline squash, the contents of the TLB \pc are cleared.

%

\begin{figure}[!t]

\includegraphics[width=\linewidth,height=0.18\textheight]{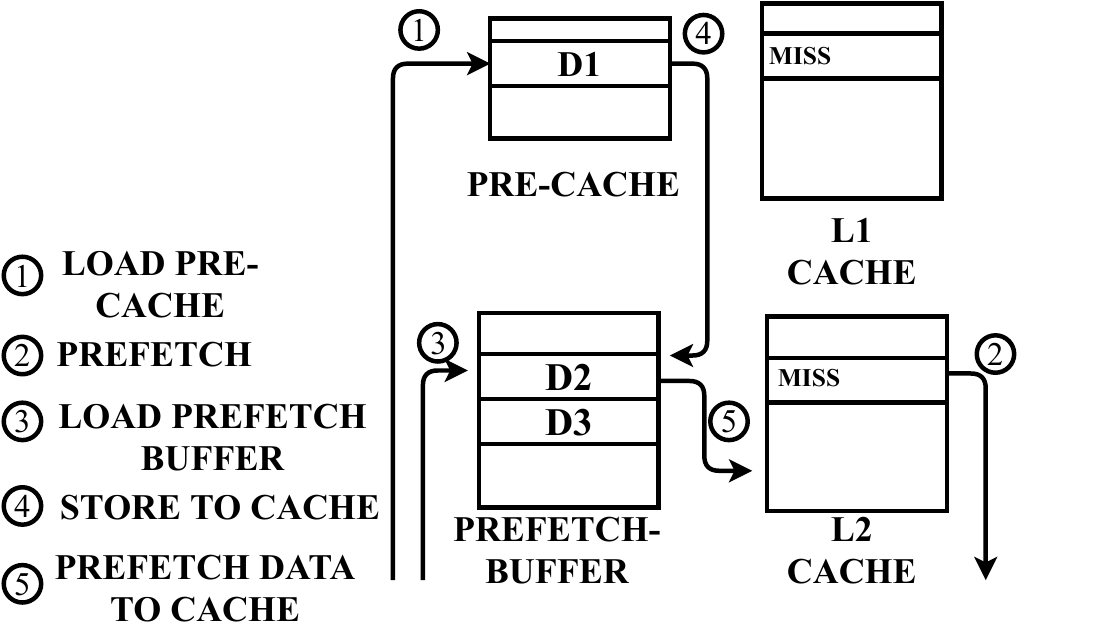}
\vspace{-18pt}
\caption{PrefetchBuffer.}
\label{fig:prefetch_buffer}

\end{figure}

\HC{Pre-fetchers can also be exploited to launch Meltdown/Spectre-style attacks. A transient load can trigger fetching of data from addresses that are related to the address of the load. Upon a squash, data pre-fetched into the cache can be exploited as a side channel. This vulnerability can be eliminated by 
 maintaining buffers at each level of the cache hierarchy. Any prefetched data at a cache level that corresponds to an uncommitted load is placed in the buffer. This data is indexed by the address of the uncommitted load. Once the load is committed, the STC of the corresponding load will be sent to the cache and the buffer at that level, triggering transfer of data from the buffer to the cache as shown in \Cref{fig:prefetch_buffer}. When the store to cache of the corresponding load arrives before the prefetched data is fetched from lower level cache, the prefetched data can be directly sent to the cache. We support this by maintaining a commit bit for each entry in the buffer to indicate whether the load corresponding to the entry has been committed. If the commit bit is set, prefetched data is directly sent to cache, otherwise it is stored in the buffer. On pipeline squash, the buffers at all the levels in cache hierarchy are cleared. Also, the buffers are designed to service any load requests sent to L2. Data that hits in a prefetch buffer is put into the pre-cache instead of L1.}

ret2spec~\cite{ret2spec} manipulates ``shared'' return stack buffers (RSBs) to trigger misspeculations that give attackers similar capabilities to the documented Spectre attacks, e.g., leaking passwords another user enters on a shared system.  It can leak sensitive information from other processes or even the same process. Since side effects of the misspeculations to memory structures are protected by \pc, ret2spec is also prevented by \pc. 

SgxPectre~\cite{SgxPectre} also exploits speculative execution to
leak secret data in an SGX enclave. Similar to the \md and \spt attacks,
SgxPectre induces the process to generate side effects of
speculatively-executed enclave code in cache. Since \pc 
catches these side effects of speculative execution, it also prevents SgxPectre.

Foreshadow~\cite{weisse2018foreshadowNG} extracts keys in SGX enclaves using a meltdown-style attack of speculatively reading data from the enclave, computing an address based on the data, and leaving side-effects in the cache by loading the newly computed address. Since \pc can catch these side-effects, it can prevent Foreshadow.

Recently, new variants -- Speculative Store Buffer Bypass~\cite{SSB}, Rogue System Register Read~\cite{RSRE}, and Lazy FPU State Restore~\cite{LFSR} -- have been proposed. These variants exploit different microarchitectural structures, such as the store buffer, system registers, and FPU registers to produce side effects in the cache that can be exploited using a Flush+Reload-style of exfiltration. Since \pc prevents cache side effects from speculative execution, these attacks are also prevented by \pc.




\noindent \textbf{Variants based on speculative stores}: 
In an architecture where a store obtains exclusive access to the block it writes before the commit stage by sending invalidations at issue, a speculative store can change the coherence state of the block in another core. This opens a side channel that can be attacked~\cite{meltdownprime}. To prevent such attacks, invalidations should be completed only after a store is non-speculative.
One way to achieve this behavior is to pre-cache invalidations by procuring a lock for a block when an invalidation is sent, preventing access to subsequent requests, and only completing the invalidation if the corresponding store commits. If the store gets squashed, the locks are released. In this method, no permanent side channel is created, and the additional delay in releasing the locks is insufficient to execute the second part of a \md or \spt attack.

Another potential side channel could be created in a cache system with a write-back write-allocate policy, when a store miss loads a block from a lower memory level. Blocks allocated and loaded by a speculative store could be used in a side channel attack. To prevent such attacks, a \pc-enabled architecture loads blocks to \pc and writes the modified block to cache using STC after the store commits.

In summary, \pc is a general solution to prevent variants of \md and \spt that exploit memory structures; no matter which buffer is exposed for an attack, \pc can prevent it by strictly containing side effects of misspeculation.

\vspace{-6pt}
\subsection{Quantifying \pc Overheads}
\vspace{-1pt}

In this section, we discuss hardware overheads in terms of implementation, area, and power. Runtime overheads are discussed in \Cref{sec:results}.
At most, the number of loads that must be buffered in the data \pc is equal to the number of in-flight loads in the pipeline. As such, it is sufficient that the number of entries in the \pc equals the number of entries in the load queue. Thus, the \pc will be significantly smaller than the cache(s) it protects in a processor. In the processor we used for evaluations (details in \Cref{sec:methodology}), the load queue has 32 entries, which corresponds to a \pc size of 2 KB -- less than 0.046\% of the  total effective cache size for a single core (\textgreater 4 MB), or roughly 6.25\% of a single L1 data cache size (32KB). Since each hardware thread in a core needs a separate \pc to ensure data isolation, the \pc should be replicated to support the number of hardware threads in a core, but the total size still need only be as large as the load queue.

Each cache level should maintain directory entries to track the addresses that
are currently in the \pc. This overhead is again negligible compared
to the total amount of cache in a core. In our design, the overhead is
512 Bytes per core. 

Similar overheads will be incurred for an instruction \pc. 
In the worst case, the instruction \pc must have capacity to hold the maximum number of in-flight instructions.
In our processor, this equals the sum of the sizes of the fetch queue, $\mu$OP
queue, and the ROB -- 448 entries. 
To propagate the instruction address along with the micro-instruction through the pipeline, an 9-bit tag is needed for every entry in fetch queue, $\mu$OP queue, and ROB, which would amount to a total of 504 B. Assuming an average instruction length of 4 B, the i\pc would need to hold 28 blocks, amounting to 1.75 kB. The directories of i\pc would need 448 B. If we use the basic block-based alternative approach, to store the basic block length queue, in the extreme worst case (448 back-to-back indirect jumps), 448 8-bit counter values would be needed, plus one additional byte to store the index of the oldest jump in the \pc. In total, the memory required to support the instruction \pc in the worst case would be roughly 6\% of the size of an L1 cache.\footnote{Note that this is a conservative figure that accounts for the worst case -- a program that fills the entire pipeline with indirect jumps that all reside in different cache blocks and straddle the block boundaries.}

Assuming, all in-flight loads point to addresses in different pages, the pre-TLB requires a maximum of 32 entires (the number of entires in the load queue). This corresponds to a pre-TLB size of 2 KB. 
For a hardware prefetcher, the pre-cache size is increased to accommodate the number of lines in the stream buffer. Typically, stream buffer size is low (e.g., 1-16), leading to a negligible increase in overhead.  RSB can be protected by maintaining a pre-cache buffer of the same size as the RSB, which also constitutes a negligible increase in overhead.

The power overhead of adding a \pc is expected to be small, since the energy consumed in accessing the \pc is significantly lower than the energy consumed in accessing the L1 cache. In our design, a \pc consumes 9$\times$ less energy than an L1 cache designed in a 32nm technology node, based on our evaluation in CACTI~\cite{cacti}.

The area overhead of adding a \pc is also expected to be small, since the number of entries in the \pc is significantly less than the cache space it protects. Also, the \pc shares most of its surrounding logic, such as access lists and coalescing logic, with L1. The data array area for the \pc in our design is 10\% of the area of the data array of our L1 cache for a 32 nm implementation. We estimated the area overhead using CACTI.

\vspace{-6pt}
\section{Implementation of \pc}
\label{sec:methodology}
\vspace{-1pt}

We implemented our proposed design on a cycle-accurate
microarchitectural simulator -- Multi2sim~\cite{multi2sim}. 
We simulated \pc on both 2-core and 4-core x86 processors with 4 and 8 hardware threads, respectively, 128-entry re-order buffer, 128-entry fetch queue, and 64-entry load-store queue. We used an inclusive cache hierarchy with private L1 and L2 caches and a shared L3. L1 instruction and data caches are 4-way, 32 KB caches, each with 2 ports and an access latency of 4 cycles. L2 is an 8-way, 2 MB cache with 4 ports and an access latency of 10 cycles. L3 is a 16-way, 8 MB cache with 8 ports and an access latency of 40 cycles. The memory network also models network latencies and queuing delays. All caches use an LRU replacement policy and 64-byte cache lines. 

The memory system of Multi2sim was modified to implement \pc and \pc directories. The load request in the memory system was restructured to load a block into \pc without affecting the levels of cache where the block was missed. Any subsequet access to the block, before the parent load commits, is modeled to be serviced with an access latency of 4 cycles. This is done to ensure that the \pc access time matches that of the L1 cache. Cache hit-level information is stored in \pc to guide STC execution. New memory requests such as STC and \pc clear were introduced to support \pc behavior in Multi2sim. The x86 architecture pipeline of Multi2sim was modified to generate the STC request on a load commit and \pc clear request on a pipeline squash. STC uses the cache hit-level information stored in \pc to move the block accessed by a committed load from \pc to cache, as explained in \Cref{sec:cache_coherence}. To ensure that the introduction of STC and \pc clear requests does not affect memory consistency, the following memory request ordering is implemented.  
Loads need not wait for STC events to finish, since they do not change the state of memory and thus do not affect consistency. Also, a pending STC is discarded if the memory location is modified by a store, since the value from the STC is stale if a store has intervened between the issue and STC operations of a load. A store removes the block from \pc before updating other levels of cache. Further, as explained in \Cref{sec:cache_coherence} we maintain \pc directory entries in the memory hierarchy and update the directories to ensure correct mirroring of the contents of the \pc at each level. A \pc clear request does not affect memory consistency, since the entries cleared from \pc do not modify the state of memory. 
Note that a clear request of \pc squashes only the blocks in \pc that were fetched by squashed load requests. To ensure that no pending or in-flight load requests affect the \pc after it is squashed, we squash all load requests in the L1 access list. However, in-flight STCs are allowed to complete, since they correspond to committed instructions. We maintain a bit for every block in \pc to indicate if the block is locked by a STC. 
A clear request of \pc clears the block only if this bit is not set. 
Since the access time difference between different levels of cache allows  blocks to be fetched into \pc before a squash request is complete, blocks belonging to load requests issued after the pipeline squash are not cleared from the \pc. To achieve this, a \pc-clear request is prioritized before any other memory request, and the squash signal clears the \pc, then proceeds to successively lower levels of cache to clear entries from \pc directories.





\vspace{-6pt}
\section {Evaluation}
\label{sec:eval}
\vspace{-1pt}

In this section, we systematically evaluate \pc. We analyze its
correctness, verify its effectiveness in preventing \md and \spt attacks, and evaluate its performance using standard benchmarks. For our evaluations, we selected benchmarks from SPEC CPU2006~\cite{spec2006}, PARSEC~\cite{parsec}, and SPLASH-2~\cite{SPLASH} suites. From these suites, we selected all the benchmarks that presented no compatibility issues with our simulator. The resulting benchmarks contain the characteristic features that are essential for testing our design. The SPEC 2006 benchmarks we used consist of equal proportions of integer and floating point benchmarks and have a working set ranging from 0.4MB to 800MB. The multi-threaded PARSEC benchmarks that we used have a working set ranging from 2MB to 64MB. PARSEC benchmarks display properties from minimal data sharing to substantial data sharing~\cite{parsec}. We ran all 11 Splash benchmarks in our evaluation. The exact benchmarks are listed in \Cref{tab:bms}.  
We fast-forwarded single-threaded applications by 1 billion instructions, while multi-threaded benchmarks were fast-forwarded to the point where threads are spawned. All benchmarks were run in detailed simulation mode for 1 billion instructions.
\begin{table}
\caption {\bf Benchmarks}
\vspace{-8pt}
\label{tab:bms}
\centering 
\small
\begin {tabular} {|l|l|}
\hline 
\multirow{ 2}{*}{\textbf{SPEC2006}} & dealII, Hmmer, H264ref, bzip2, Calculix\\ 
& Astar, omntepp, GemsFDTD, Povray \\ \hline 
\multirow{ 2}{*}{\textbf{PARSEC}} & blackscholes, fluidanimate, streamcluster, \\
& swaptions \\ \hline
\multirow{ 2}{*}{\textbf{SPLASH-2}} & fft, lu, ocean, radiosity, raytrace, radix, barnes\\
& cholesky, fmm, water-nsquared, water-spatial \\ \hline
\end{tabular}
\vspace{-16pt}
\end {table}

\vspace{-6pt}
\subsection{Correctness Analysis}
\label{sec:correctness}
\vspace{-1pt}

Pre-cache introduces a cache module that buffers data retrieved from memory and sends the data to the cache hierarchy if the data-requesting instruction is determined to be valid. \pc does not change other behaviors in the processor. As discussed in \Cref{sec:cache_coherence}, we carefully design \pc to ensure cache coherence and memory consistency in a multi-core processor and maintain inclusivity in a multi-level cache. By design, \pc does not cause correctness issues in processors. 

One way to verify correctness of our design is to compare the sequence of loads and stores committed by the processor before and after introducing \pc. 
To verify correctness, we dumped the sequence of loads and stores committed by the processor in the baseline and \pc designs.
From the dumps, we observed that the sequence of committed load and store addresses in the \pc-based design matched the sequence in the baseline design, affirming that introduction of \pc does not affect design correctness.

To demonstrate that the hardware changes introduced by the \pc do not affect the correctness of execution, we compare the benchmark outputs from our simulator in both \pc and baseline designs for single-threaded benchmarks from SPEC CPU 2006 and multi-threaded benchmarks from SPLASH-2 and PARSEC-3.0. We found no difference in the benchmark outputs. When an application is run, results are updated only after instructions get committed. A \pc-enhanced design does not affect the instruction commit order. We compared the instruction commit order in both \pc and baseline designs and observed that the order did not change, confirming that the \pc does not affect the correctness of execution of a program.

\vspace{-6pt}
\subsection{Security Analysis and Evaluation}
\vspace{-1pt}

In this section, we systematically evaluate the security and the effectiveness of \pc in preventing the \md and \spt attacks through a three-pronged security analysis.

\vspace{-4pt}
\begin{enumerate}[leftmargin=*,noitemsep]
\item We analyze the potential attacks against \pc and discuss how \pc prevents the attacks.
\item We
systematically track the effects of transient instructions and show that the effects are strictly contained by \pc, confirming that the effects are not leaked to attackers.
\item We evaluate the effectiveness of \pc based on the \md and \spt attacks. We conservatively simulate the attacks and show that \pc can defeat them.
\end{enumerate}

\vspace{-6pt}
\subsubsection{Potential attacks}
\vspace{-1pt}

There are three potential security hazards in \pc: (1) adding additional 
``buffers'' in the processors may have side effects; (2) timing attack
against \pc; (3) side-channels that might be opened due to the presence of \pc. 
To be effective, \pc must contain all side effects in 
various processor buffers. In our current implementation, we have
shown how to pre-cache data and instructions; this  prevents
all existing variants of \md and \spt, as well as new variants we presented in \Cref{sec:iattacks}. However, additional
microarchitectural buffers also have side effects that should be contained
by \pc.  
\pc's approach is general. If a new buffer is introduced in a
processor, \pc can be extended at the same time to cover it. We do not foresee
fundamental challenges in designing these extensions.

A reader may wonder, since \pc acts like a cache, will introduction of \pc lead to new attack variants that target \pc? Such timing attacks are impossible in practice, since the data exits the \pc before an attacker has a chance to access it. A timing attack on \pc would require the fetched secret data to leave an observable side effect. Since the \pc is flushed after the pipeline squash caused by the access of secret data, thereby leaving no observable side effects, such timing attacks are not possible. 
Even if an attacker tried to infer the secret value (step 4 in \Cref{fig:attack-abstract}) by using a timing-based attack such as Flush+Reload~\cite{flush+reload} on the \pc, before a pipline squash, using another hardware thread on the same core, since we isolate data between hardware threads in the \pc by either tagging the data with the thread ID or by maintaining as many Pre-caches as the number of hardware threads, this attack is also impossible.

 To consider side channels that might be opened by the presence of \pc, we must understand the effects of \pc on the state of the cache. Since \pc buffers data accessed by all loads that missed in L1 until the loads are committed, \pc has the effect of delaying the corresponding blocks from entering L1. This could prioritize a missed block in L1 over a block that was hit in L1, because the LRU ordering for the missed block is updated at the time of instruction commit rather than the time of access. While this might seem like a potential side-channel, note that only blocks accessed by valid loads are allowed to enter L1. The blocks accessed by invalid loads are flushed from \pc during a pipeline squash. Thus, invalid loads have no impact on LRU state, as LRU state contains no information about invalid loads.
Consequently, no information about invalid loads can be obtained by observing any changes in LRU state caused by the presence of \pc. Our benchmark evaluations also show that any changes in LRU state due to the presence of \pc do not affect cache hit rate, compared to the baseline archiecture.

\vspace{-6pt}
\subsubsection{Preventing the effects of transient instructions}
\vspace{-1pt}

To show that \pc prevents transient instructions from affecting a processor's cache, we perform evaluations involving two types of transient instructions. The first type reads data from a privileged memory location. The second type is a speculatively-executed load caused by a branch misprediction.
Every memory load in the processor is committed only after privilege check passes. This is generally performed in the last stage of the pipeline. If the privilege check fails, an exception is raised and the pipeline is squashed. However, in a traditional architecture, data fetched by the violating instruction is already allocated in the cache, leaving a side effect of the instruction. To simulate this behavior, we mark a certain address range as privileged in our simulator. The simulator is modified to raise an exception at commit if the address corresponding to a load is marked as privileged. We then run a program with a privileged load instruction on the baseline and \pc designs. The privileged load causes an exception to be raised at the commit stage, and the pipeline is squashed. After the pipeline squash, we observe the contents of the cache in both designs. In the baseline design, the cache contained the data from the privileged memory location, while in the \pc design the data was not in the cache. \pc prevents an illegal access to a privileged address from loading data to the cache, eliminating cache side effects.

A load will also be squashed if it belongs to a mispredicted execution path. Resolution of misprediction involves squashing all the instructions in the pipeline. However, the load of interest could have placed data in the cache before the squash, leading to side effects. To simulate this behavior and evaluate the effectiveness of \pc in preventing such side effects, we modify our simulator to tag every address that is speculatively loaded. Once the load corresponding to the address is resolved to be non-speculative, the address is untagged. We then run a program that suffers a branch misprediction (a simple for loop can achieve this). Upon pipeline squash due to misprediction, we terminate the simulation and dump the cache contents. From the cache dumps of the baseline and \pc designs, we observed that the cache contained tagged addresses in the baseline design but not in the \pc design, confirming that \pc eliminates side effects of speculative execution in the cache.

In both evaluations described above, after a pipeline squash, we wait for any in-flight memory accesses to finish before dumping the cache. Since an in-flight access, after the squash, could affect the contents of the cache, premature dumping of the cache could compromise the validity of our evaluations.

\vspace{-6pt}
\subsubsection{Preventing \md and \spt attacks}
\vspace{-1pt}

We evaluated the effectiveness of \pc in preventing Meltdown and Spectre by running the first part of the actual attacks on the baseline and \pc designs. The second part of the attacks (Flush + Reload~\cite{flush+reload}), which observes side effects in the cache is unnecessary, since we can observe the contents of the cache directly to determine if any side effects are present. 

\noindent \textbf{Meltdown}: The first part of a Meltdown attack involves reading data from a privileged memory location and using the data to generate a legal address. If a read request to the generated memory address can be sent before the permission check of the first read fails, a side effect can be created in the cache. This side effect can be exploited in the second part of the attack using a cache-timing attack such as Flush+Reload~\cite{flush+reload}. We simulated the first part of the attack on the baseline and \pc designs. \Cref{lst:meltdownsim} presents the attack code.

\vspace{-6pt}
\begin{lstlisting} [caption=Instruction sequence to simulate Meltdown, label=lst:meltdownsim, captionpos=t,]
;ecx = marked address
;ebx = probe array
mov al, byte [ecx]
shl eax, 0xc
mov ebx, qword [ebx + eax]
\end{lstlisting}

We mark a group of addresses in our simulator to be inaccessible to the program and modify the simulator to raise an exception in the commit stage if any of these addresses is accessed. We add a delay to the exception handling to ensure enough time for the subsequent instructions to finish execution. We then execute the code in \Cref{lst:meltdownsim}. Below, we explain the three instructions in the code.

\noindent $\bullet$ The instruction at line 3 reads data from a marked address into a register.

\noindent $\bullet$ The instruction at line 4 uses the value in the register to generate an offset. 

\noindent $\bullet$  The instruction at line 5 uses the offset to generate an address to which a memory read is issued, creating a cache side effect.

Since the instruction at line 3 reads data from a marked address, our simulator raises an exception and squashes the pipeline. We dump the contents of the cache after all outstanding memory requests complete. In the baseline design, the cache contained the address generated by the instruction at line 5; the attack was successful. In the \pc design, the cache dump did not contain the address generated by the instruction at line 5; the attack was prevented. In fact, even the marked address from the instruction at line 3 was not in the cache of the \pc design, while the baseline design's cache contained the marked address, further confirming \pc effectiveness in preventing \md.








\noindent \textbf{Spectre}: The first part of a Spectre attack involves training a branch predictor to misspeculate and cause a branch to be speculatively executed. A speculative instruction following the branch loads data from memory locations that the application is not allowed to access. After speculation is resolved, the instructions in the misspeculated branch path are squashed. However, data that has been speculatively loaded remains in the cache. This side effect is exploited by the second part of the attack, using a technique such as Flush+Reload~\cite{flush+reload}. We simulated the first part of the attack on the baseline and \pc  designs. \Cref{lst:spectresim} presents the attack code, consisting of a \texttt{for} loop with a simple body with no branches. The code is designed based on the fact that when the loop iterator reaches the loop bound, the branch predictor, which has been trained to take the branch and execute another iteration of the loop, mispredicts the control flow towards the body of the loop instead of outside the loop. Below we explain the code in detail.

\noindent $\bullet$  Line 3 shows the setup of the \texttt{for} loop. The register \texttt{eax} holds the value of the iterator, \texttt{i}.

\noindent $\bullet$  The instruction at line 5 copies the iterator to another register, \texttt{ecx}.

\noindent $\bullet$  The instruction at line 6 uses the value in the register \texttt{ecx}  to generate an offset. 

\noindent $\bullet$  The instruction at line 7 uses the offset to generate an address to which a memory read is issued, creating a cache side effect.


\vspace{-6pt}
\begin{lstlisting} [caption=Instruction sequence to simulate Spectre, label=lst:spectresim, captionpos=t,]
//eax = iteration index
//ebx = probe array
for (i = 0; i < 100; i++) 
{
  _asm_ { mov ecx, eax
          shl ecx, 0xc
          mov ebx, qword [ebx + ecx] };
}
\end{lstlisting}

Since the code is designed to cause a branch misprediction, instructions in lines 5-7 are executed \textit{transiently} after the last loop iteration. This means that the instruction in line 7 can read data from an address outside the bounds of the application, potentially from another application's memory space. While the instructions will be squashed after the branch prediction is resolved, line 7 causes a cache side effect.

We ran the code of \Cref{lst:spectresim} on the baseline and on \pc designs, and dumped the cache contents after all outstanding memory requests completed, just after the pipeline squash. We observed that the cache dump of the baseline design indeed contained the address generated in line 7; however, the cache dump of the \pc design did not contain the address, confirming that \pc prevents Spectre attacks. 




\vspace{-6pt}
\subsection{Performance Evaluation}
\label{sec:results}
\vspace{-1pt}

In this section, we analyze the impact of \pc on the
performance of several benchmarks from the SPEC CPU2006, PARSEC, and SPLASH-2 benchmark
suites. Our goal is to show that our \pc microarchitecture achieves the security goals we set forth without imposing a significant performance overhead. 

\begin{figure*}[!t]
\includegraphics[width=\linewidth,height=0.18\textheight]{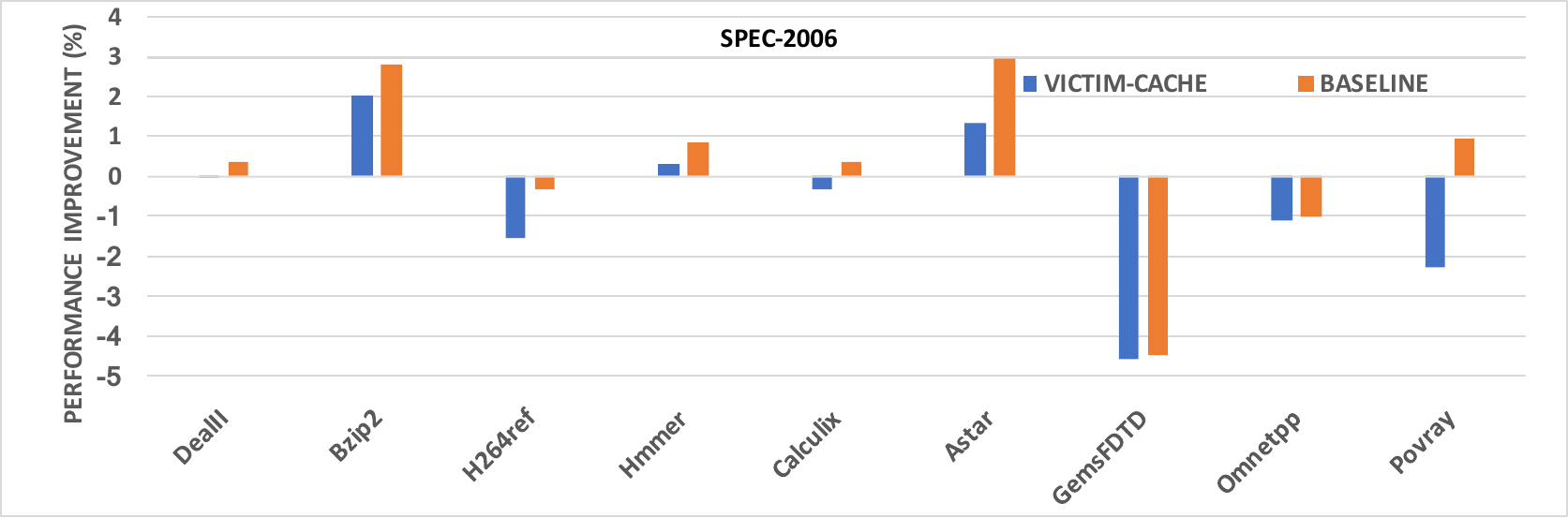}
\vspace{-18pt}
\caption{Performance impact of Pre-cache over the baseline design and a design with a victim cache in a single-core design.}
\label{fig:perf_results}



\includegraphics[width=\linewidth,height=0.18\textheight]{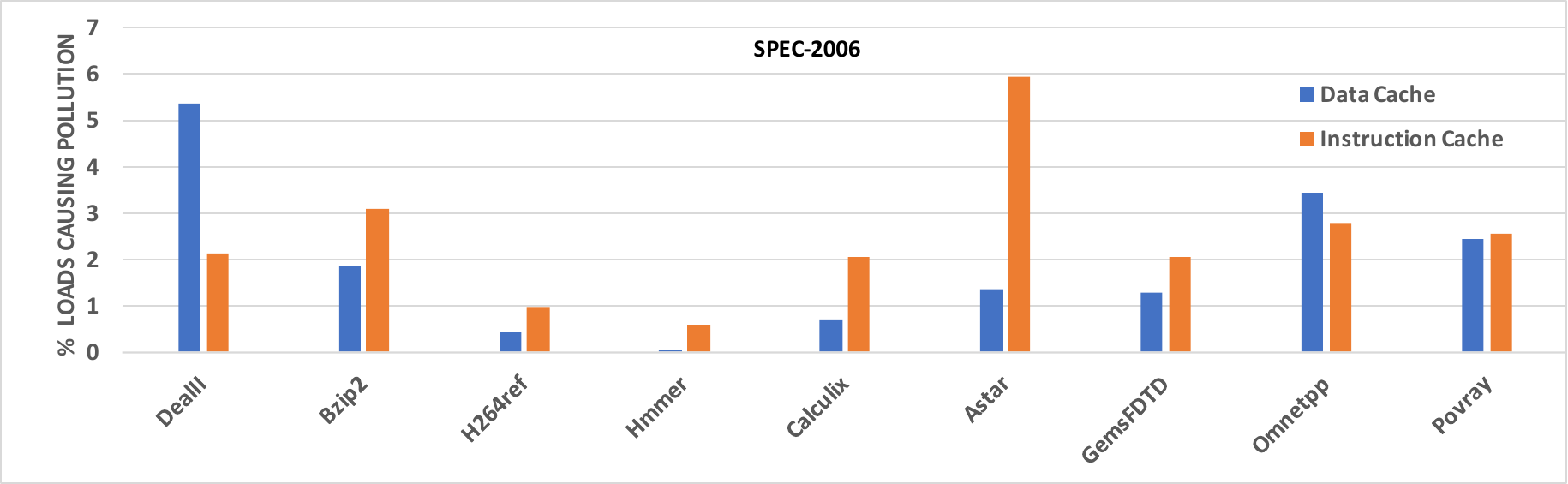}
\vspace{-18pt}
\caption {Percentage of loads causing cache pollution in the L1 instruction and data caches in a single core design.}
\label{fig:polluting_loads}

\includegraphics[width=\linewidth,height=0.18\textheight]{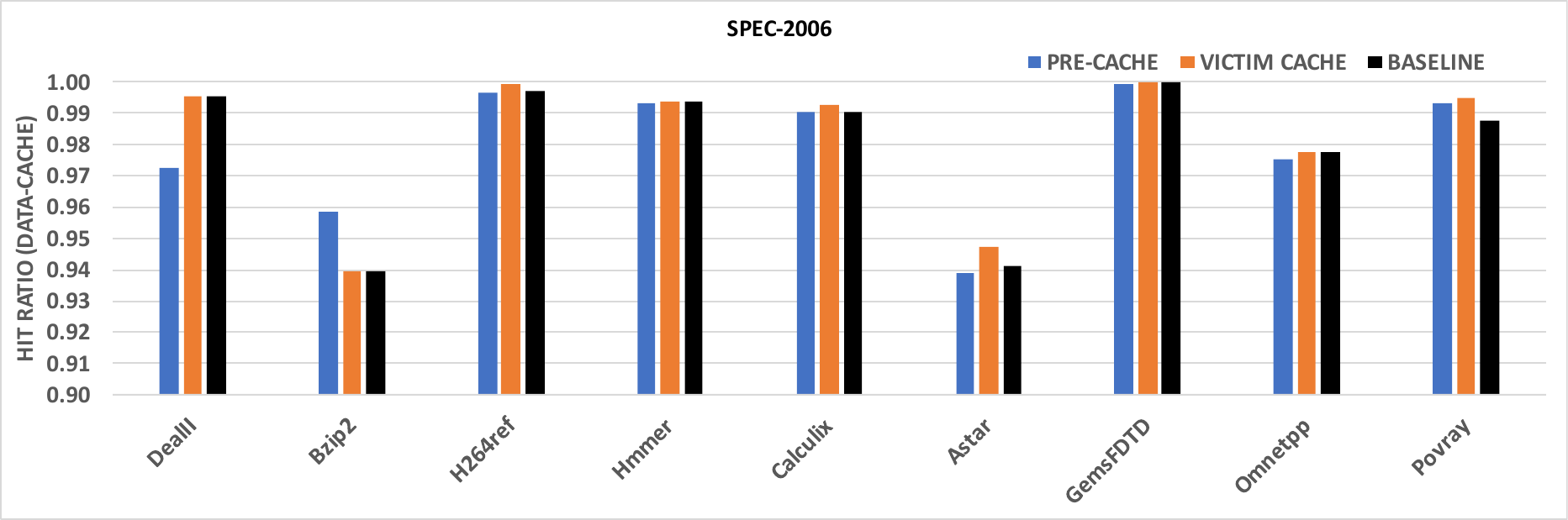}
\vspace{-18pt}
\caption{Cache hit rate for the L1 data cache subsytem in a single core design.}
\label{fig:Hit_rate_results_D}

\includegraphics[width=\linewidth,height=0.18\textheight]{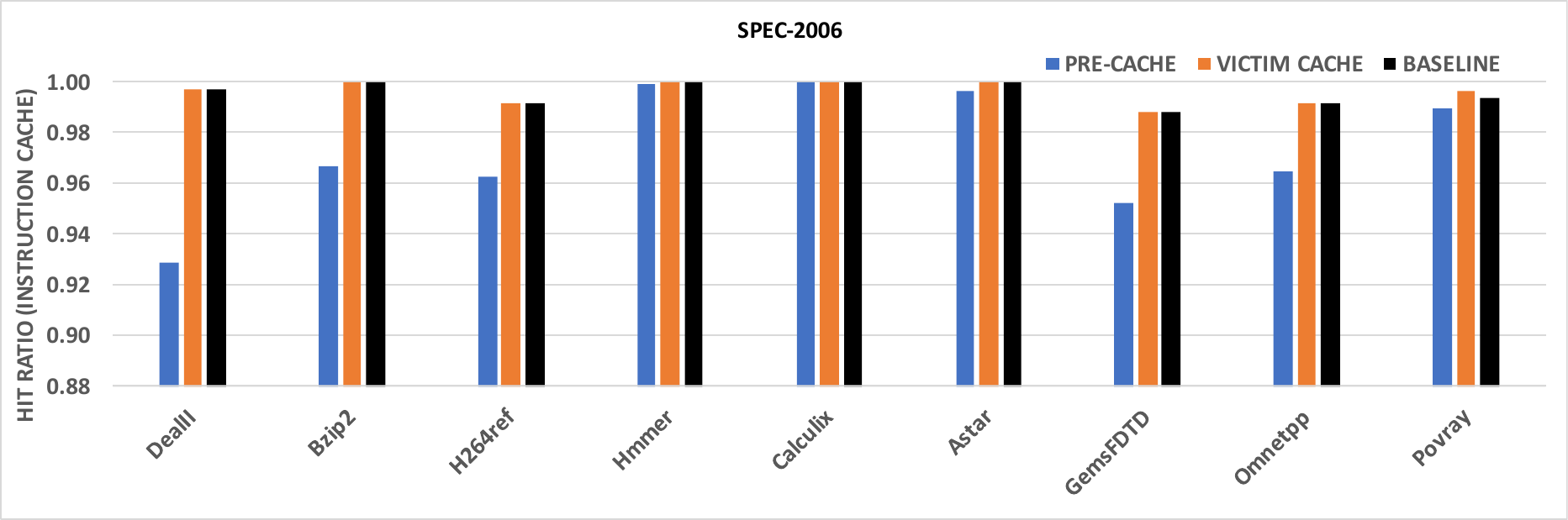}
\label{fig:Hit_rate_results_I}
\vspace{-18pt}
\caption{Cache hit rate for the L1 instruction cache subsytem in a single core design.}



\vspace{-14pt}
\end{figure*}

We compare the performance of
our \pc microarchitecture against two baseline designs. One is the baseline microarchitecture we started with before making changes to support \pc. 
Since \pc slightly increases the effective size of the L1 cache, we also compare against a second baseline design that includes a victim cache~\cite{jouppi1998improving} for the
L1 cache. The size of the victim cache is
equal to the size of the \pc -- 32 cache blocks, which amounts
to 2KB. \Cref{fig:perf_results} presents the impact of \pc on the
performance of the processor, measured as committed instructions per
cycle (IPC). For each benchmark, we show performance improvement of the \pc microarchitecture over each baseline design. 


\begin{figure*}[!t]
\includegraphics[width=\linewidth,height=0.18\textheight]{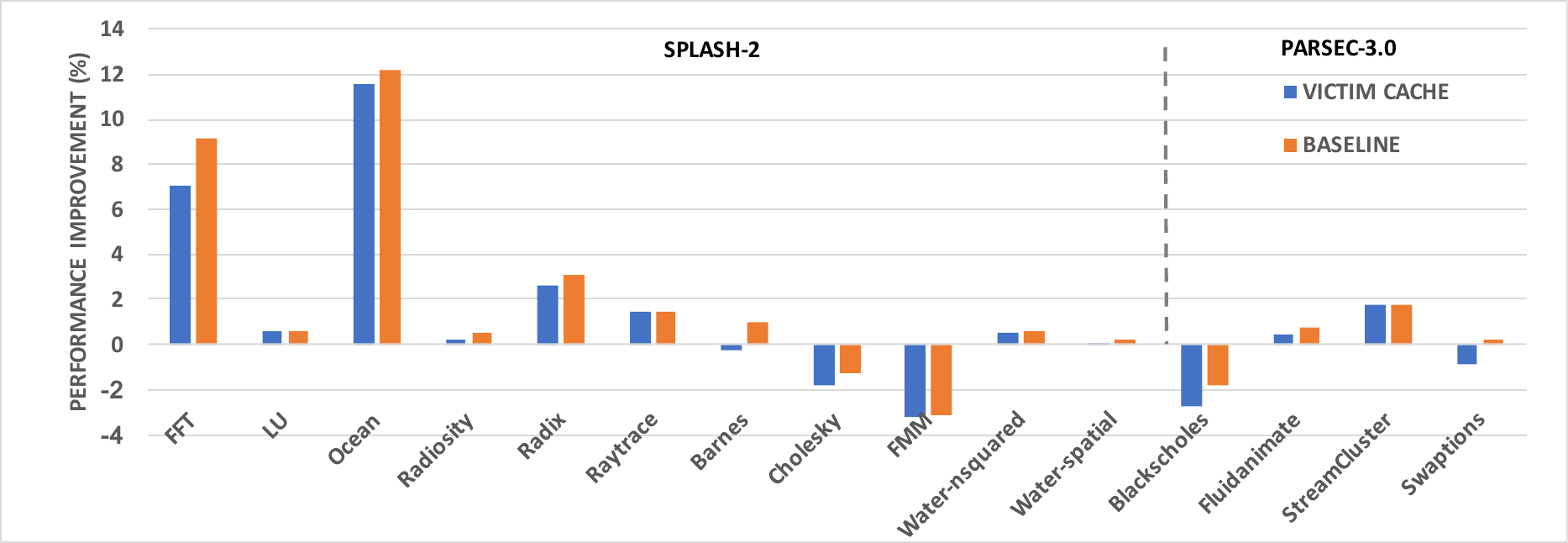}
\vspace{-18pt}
\caption{Performance impact of Pre-cache over the baseline design and a design with a victim cache in 2-core design.}
\label{fig:2_core_perf_results}

\includegraphics[width=\linewidth,height=0.18\textheight]{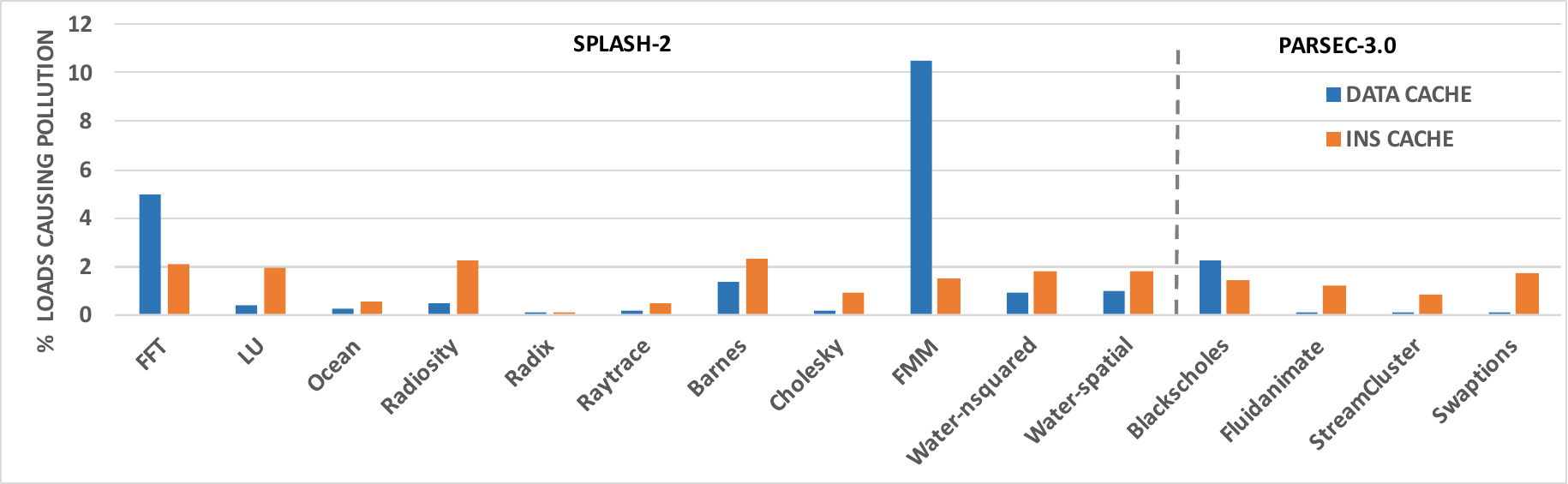}
\vspace{-18pt}
\caption {Percentage of loads causing cache pollution in the L1 instruction and data caches in a 2-core design.}
\label{fig:2_core_polluting_loads}

\includegraphics[width=\linewidth,height=0.18\textheight]{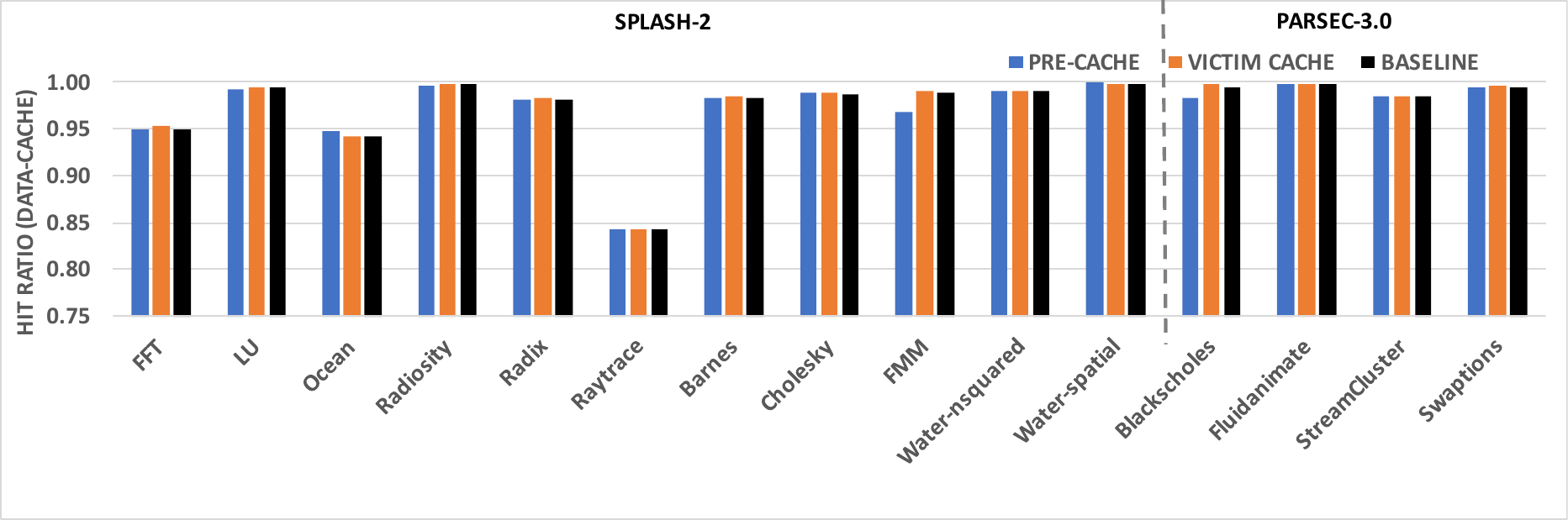}
\vspace{-18pt}
\caption{Cache hit rate for the L1 data cache subsytem in a 2-core design.}
\label{fig:2_core_Hit_rate_results_D}
\includegraphics[width=\linewidth,height=0.18\textheight]{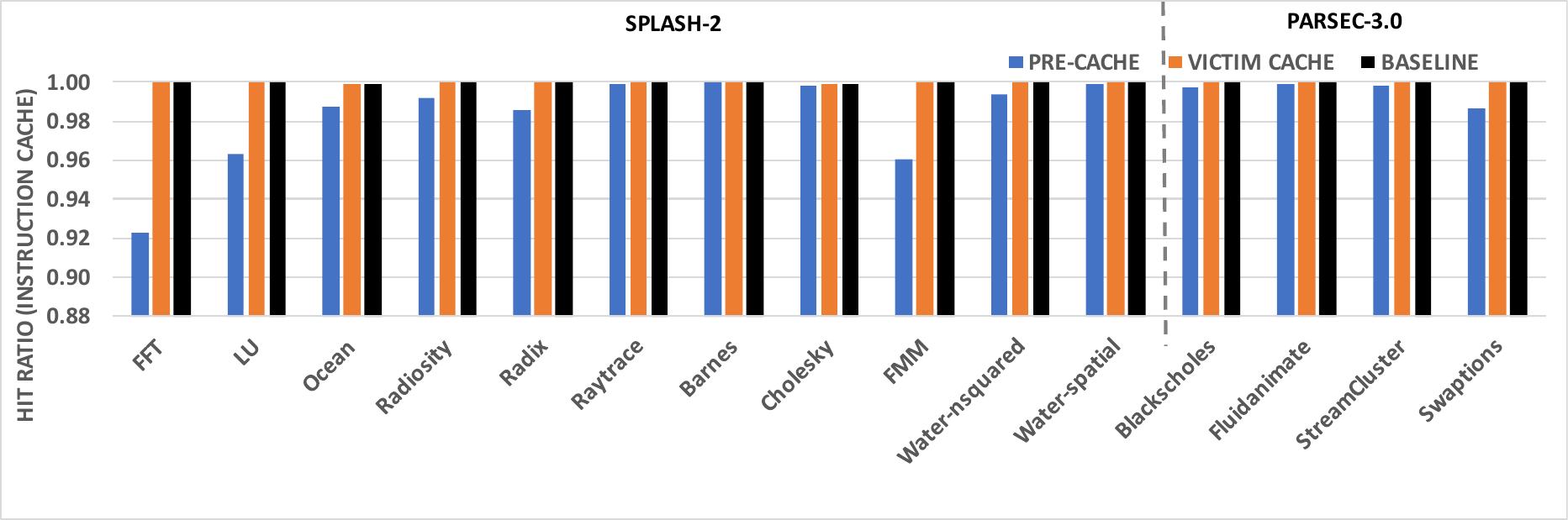}
\vspace{-18pt}
\caption{Cache hit rate for the L1 instruction cache subsytem in a 2-core design.}
\label{fig:2_core_Hit_rate_results_I}
\end{figure*}
\begin{figure*}[!t]
\includegraphics[width=\linewidth,height=0.18\textheight]{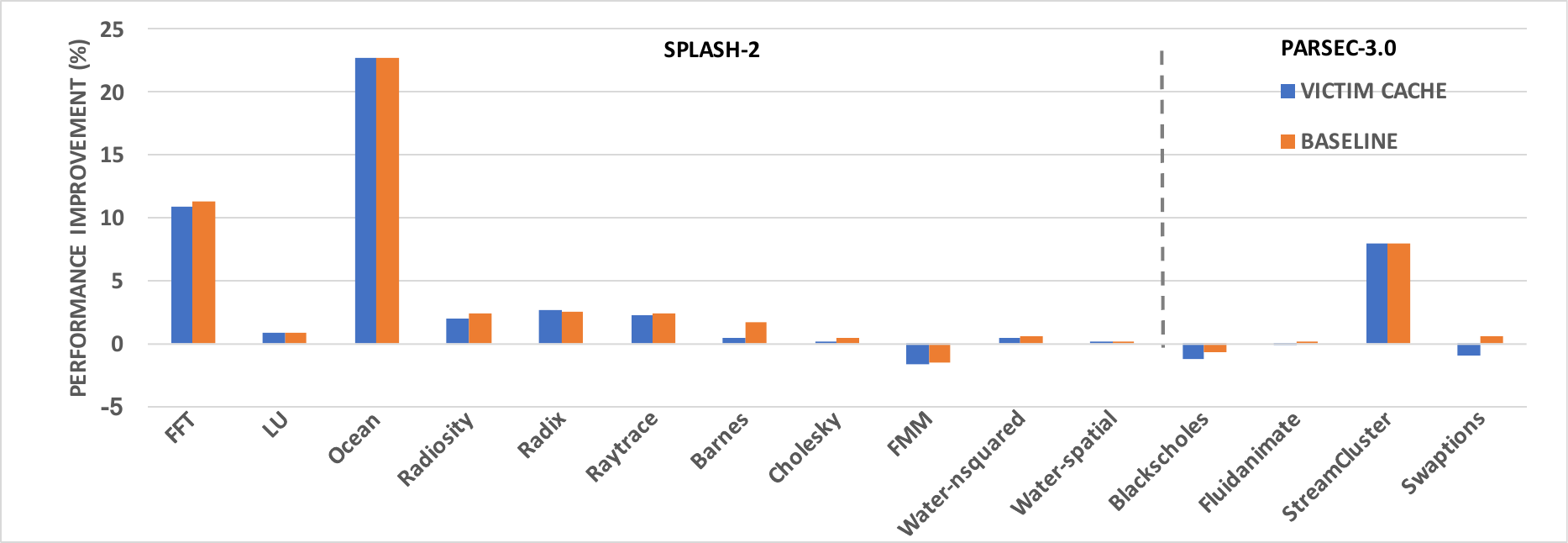}
\vspace{-18pt}
\caption{Performance impact of Pre-cache over the baseline design and a design with a victim cache in 4-core design.}
\label{fig:4_core_perf_results}
\includegraphics[width=\linewidth,height=0.18\textheight]{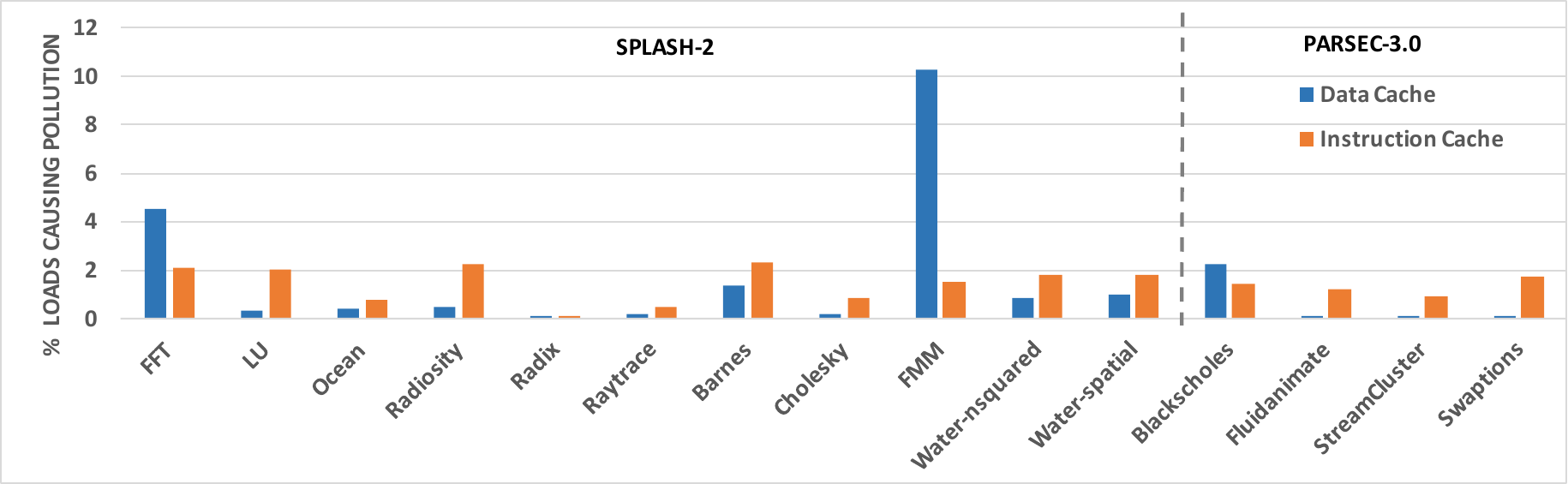}
\vspace{-18pt}
\caption{Percentage of loads causing cache pollution in the L1 instruction and data caches in a 4-core design.}
\label{fig:4_core_polluting_loads}
\includegraphics[width=\linewidth,height=0.18\textheight]{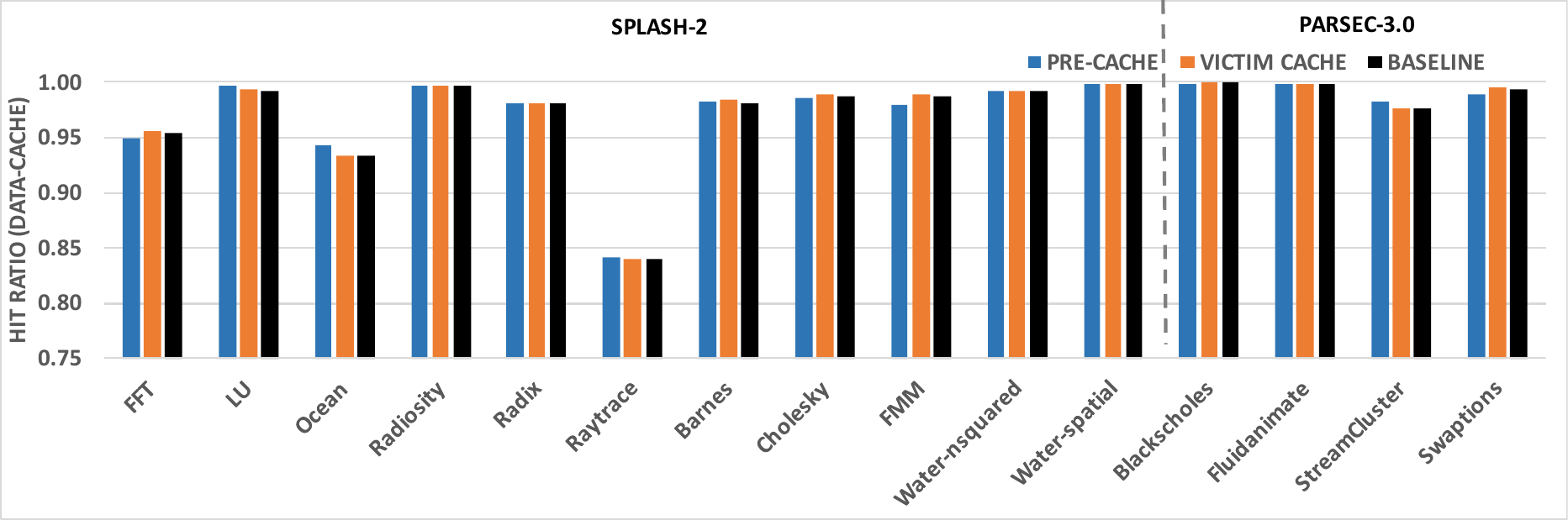}
\vspace{-18pt}
\caption{Cache hit rate for the L1 data cache subsytem in a 4-core design.}
\label{fig:4_core_Hit_rate_results_D}
\includegraphics[width=\linewidth,height=0.18\textheight]{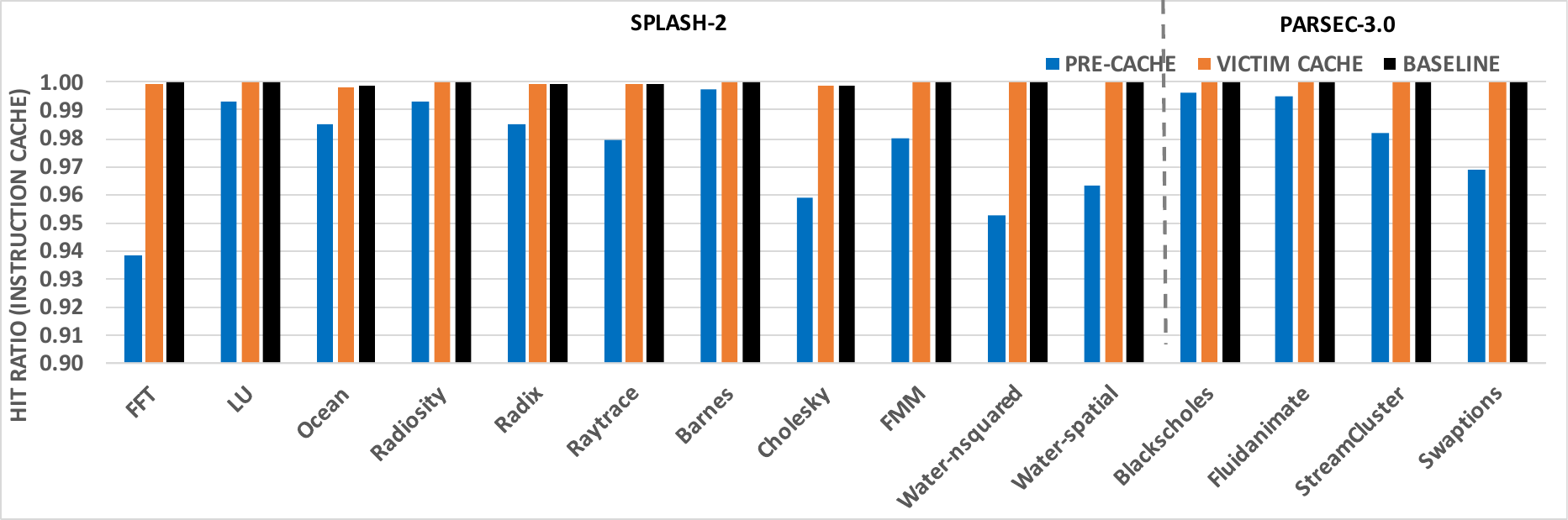}
\vspace{-18pt}
\caption{Cache hit rate for the L1 instruction cache subsytem in a 4-core design.}
\label{fig:4_core_Hit_rate_results_I}
\end{figure*}



%
%

The results in \Cref{fig:perf_results}, \Cref{fig:2_core_perf_results}, \Cref{fig:4_core_perf_results} show the performance improvement of a \pc enhanced design over baseline and victim cache designs for benchmarks running on single core, two core and four core designs, respectively. The figures show that for majority of the benchmarks, \pc improves performance or has a negligible performance impact compared to the baseline. Performance improvement compared to the original baseline design can be attributed to the slightly larger effective cache size of the \pc-augmented design and the fact that \pc prevents cache pollution caused by transient instructions.
\Cref{fig:polluting_loads} shows a metric for cache pollution -- the percentage of squashed loads that result in a cache eviction. \Cref{fig:2_core_polluting_loads} and \Cref{fig:4_core_polluting_loads} shows the same metric for a two core and a four core design respectively.
While the presence of a victim cache reduces the impact of cache pollution, the conventional microarchitecture with victim cache still allows cache pollution. For example, when a speculatively-fetched block (pollutant) is put in a cache line, it causes an eviction. The evicted block is moved to the victim cache. When the evicted block is read again, it is put back into the cache. However, the pollutant is still in the cache (or the victim cache) if it was not the least-recently-used cache block. 
Thus, the victim cache can mitigate the impact of cache pollution only to a limited extent, but it does not prevent pollution caused by speculative loads. \pc, however, avoids cache pollution caused by speculative loads completely. 


\pc improves performance for some applications more than others.
For example, Ocean and FFT both partition a matrix dataset into sub-matrices, assigned to different threads.
Misspeculation causes a thread assigned to a sub-matrix to update elements belonging to a neighboring thread at the boundaries between sub-matrices. Each encroachment 
causes invalidations when a thread updates a block belonging to a neighbor and when the neighbor brings the block back to its own cache. 
FFT partitions a matrix into groups of rows, so encroachment happens primarily at the end of a partition, but Ocean partitions into square sub-grids, resulting in encroachment at the end of every row. This is why \pc, which prevents cache pollution caused by encroachments, improves performance more for Ocean.
As parallelism increases, more boundaries exist between partitions, and the pollution reduction benefit of \pc increases. E.g., doubling the number of cores from 2 to 4 increases Ocean's performance benefit from 12\% to 22.7\%. 


The performance results also show that \pc degrades performance for a few benchmarks. Increased traffic in the cache network due to STC operations can contribute to performance degradation. 
%
Another reason for performance degradation is that cache pollution 
can actually improve performance for some applications. 
For example, a load that is speculatively executed before a store to the same address will eventually be squashed. However, the store and the load (re-executed in correct order) will be performed faster if the speculatively-loaded cache block remains in the cache. By preventing cache pollution, the \pc-augmented design must go to  a lower level cache or the memory to fetch the block for the store.

The benchmark with the highest performance degradation (\PerfImprvMin\%) is GemsFDTD. Even though its L1 hit rate is high, GemsFDTD has the highest number of aggregate misses per instruction among all our benchmarks~\cite{jaleel2010memory}. A large number of misses results in a large number of STCs, which increases network traffic. 
Also, GemsFDTD is primarily a streaming benchmark with a high percentage of unused evictions -- i.e., a significant number of the evicted lines are never reused~\cite{jaleel2010memory}. Preventing cache pollution provides little benefit for such an application, because the cached data have a low chance of being reused.

For the benchmark Cholesky, increasing the number of cores from two to four in a \pc design improves performance compared to the baseline. This can be attributed to a significant fraction of Cholesky's execution time spent performing synchromization~\cite{SPLASH}. The additional time required to perform synchronization for a larger number of cores provides a \pc-augmented design more time to accomodate STC traffic. Thus, the performance impact of STC traffic can be absorbed into the delays caused by synchronization.

Similarly, FMM sees a better performance benefit compared to the baseline in a four core \pc design compared to a two core design. Like GemsFDTD, FMM benefits from data loaded into the cache by squashed loads. In the four core design, the squashed loads are distributed among more cores,  reducing the time required to reload the data for each core, thereby improving the performance.

The results of Cholesky and FMM demonstrate that performance limitations caused by the addition of \pc in the form of STC traffic or re-loading data from squashed loads is reduced when the number of cores increases. To verify this across all benchmarks, we computed the average IPC for two core baseline design (2.178), four core baseline design (4.056), two core \pc-augmented design (2.203) and four core \pc-augmented design (4.167). We observed that the performance improvement for a \pc-augmented design with respect to the baseline is 3\% higher when the number of cores is increased from two to four.

\Cref{fig:perf_results}, \Cref{fig:2_core_perf_results}, and \Cref{fig:4_core_perf_results} show that addition of \pc impacts some applications more than others. This is because applications with higher cache pressure have higher sensitivity to cache pollution, and thus higher sensitivity to addition of a \pc.  
\Cref{fig:Hit_rate_results_D}, \Cref{fig:2_core_Hit_rate_results_D} and \Cref{fig:4_core_Hit_rate_results_D} show the L1 data cache hit rate for a single core, two core and four core design respectively.
Lower hit rates can indicate higher sensitivity to cache pollution (e.g., FFT, Ocean), since useful data evicted by cache pollution cause cache misses.
Pre-cache prevents valid data from being evicted by transient data for such applications, which may result in performance improvement.
In cases where cache pollution is beneficial, Pre-cache can reduce hit rate and degrade performance.
%

We also analyzed the impact of \pc on L1 instruction cache hit rate.  \Cref{fig:Hit_rate_results_I}, \Cref{fig:2_core_Hit_rate_results_I} and \Cref{fig:4_core_Hit_rate_results_I} show the L1 instruction cache hit rate for a single core, 2 core and 4 core design respectively. 
Adding an i\pc reduces the hit rate of the L1-iCache sub-system by 2\%, on average, for our benchmarks. 
This is because misspeculated instructions are likely to be valid at some point during program execution, such that misspeculation may essentially prefetch instructions in some scenarios.
Also, code size is often significantly smaller than data size for an application, making it easier to effectively cache instructions, in spite of any cache pollution.

For a four core processor, the maximum performance benefit and degradation observed from \pc are \PerfImprvMax\% and \PerfImprvMin\%, respectively, while the average performance improvement is \PerfImprvAvg\%. For a two core processor, the maximum performance benefit and degradation are 12.2\% and -3.1\%, respectively, while the average performance improvement is 1.7\%.

\vspace{-6pt}
\section{Conclusion}
\label{sec:conclusion}
\vspace{-1pt}

In this paper, we presented \pc as a microarchitecture solution to
prevent variants of \md, \spt that use memory structures in the processor as an exfiltration channel. As opposed to current widely adopted solutions 
that only target symptoms of a vulnerable processor, our solution 
addresses the security vulnerabilities exploited by the attacks at a microarchitecture level
and lays out a fundamental rule that security-conscious
designers must consider when evaluating  future microarchitectural
enhancements. \pc is a general solution that can be applied to all
current and future attack variants that use a memory-based structure as a side channel, and it does not significantly impact performance for most applications. Our solution also does not impose significant verification overheads and can be applied to multi-core systems with multi-level caches.




\bibliographystyle{IEEEtran}
\bibliography{refs}

@inproceedings{parsec,
 author = {Bienia, Christian and Kumar, Sanjeev and Singh, Jaswinder Pal and Li, Kai},
 title = {The PARSEC Benchmark Suite: Characterization and Architectural Implications},
 booktitle = {Proceedings of the 17th International Conference on Parallel Architectures and Compilation Techniques},
 series = {PACT '08},
 year = {2008},
 isbn = {978-1-60558-282-5},
 location = {Toronto, Ontario, Canada},
 pages = {72--81},
 numpages = {10},
 url = {http://doi.acm.org/10.1145/1454115.1454128},
 doi = {10.1145/1454115.1454128},
 acmid = {1454128},
 publisher = {ACM},
 address = {New York, NY, USA},
}

@inproceedings{invisispec,
  title={InvisiSpec: Making Speculative Execution Invisible in the Cache Hierarchy},
  author={Yan, Mengjia and Choi, Jiho and Skarlatos, Dimitrios and Morrison, Adam and Fletcher, Christopher W and Torrellas, Josep}
}

@article{safespec,
  title={SafeSpec: Banishing the Spectre of a Meltdown with Leakage-Free Speculation},
  author={Khasawneh, Khaled N and Koruyeh, Esmaeil Mohammadian and Song, Chengyu and Evtyushkin, Dmitry and Ponomarev, Dmitry and Abu-Ghazaleh, Nael},
  journal={arXiv preprint arXiv:1806.05179},
  year={2018}
}

@article{jaleel2010memory,
  title={Memory characterization of workloads using instrumentation-driven simulation},
  author={Jaleel, Aamer},
  journal={Web Copy: http://www. glue. umd. edu/ajaleel/workload},
  year={2010}
}

@book{Hennessy,
 author = {Hennessy, John L. and Patterson, David A.},
 title = {Computer Architecture, Fifth Edition: A Quantitative Approach},
 year = {2011},
 isbn = {012383872X, 9780123838728},
 edition = {5th},
 publisher = {Morgan Kaufmann Publishers Inc.},
 address = {San Francisco, CA, USA},
}

@inproceedings{spec2006,
 author = {J.L.Henning},
 title = {{SPEC CPU2006 Benchmark Descriptions}},
 booktitle = {ACM SIGARCH newsletter, Computer Architecture News,},
 volume = {34, No. 4},
 year = {2006},
 month = {September},
}

@inproceedings{gotzfried2017cache,
  title={Cache attacks on Intel SGX},
  author={G{\"o}tzfried, Johannes and Eckert, Moritz and Schinzel, Sebastian and M{\"u}ller, Tilo},
  booktitle={Proceedings of the 10th European Workshop on Systems Security},
  pages={2},
  year={2017},
  organization={ACM}
}

@INPROCEEDINGS{cacti, 
author={N. Muralimanohar and R. Balasubramonian and N. Jouppi}, 
booktitle={40th Annual IEEE/ACM International Symposium on Microarchitecture (MICRO 2007)}, 
title={Optimizing NUCA Organizations and Wiring Alternatives for Large Caches with CACTI 6.0}, 
year={2007}, 
volume={}, 
number={}, 
pages={3-14}, 
doi={10.1109/MICRO.2007.33}, 
ISSN={1072-4451}, 
month={Dec},}

@article{meltdown,
	author = {Lipp, Moritz and Schwarz, Michael and Gruss, Daniel and
		 Prescher, Thomas and Haas, Werner and Mangard, Stefan and
			 Kocher, Paul and Genkin, Daniel and Yarom, Yuval and Hamburg,
		 Mike},
	title = {Meltdown},
	journal = {ArXiv e-prints},
	archivePrefix = "arXiv",
	eprint = {1801.01207},
	year = 2018,
	month = jan,
}

@article{spectre,
 author = {Kocher, Paul and Genkin, Daniel and Gruss, Daniel and Haas, Werner and Hamburg, Mike and Lipp, Moritz and Mangard, Stefan and Prescher, Thomas and Schwarz, Michael and Yarom, Yuval},
 title = {Spectre Attacks: Exploiting Speculative Execution},
 journal = {ArXiv e-prints},
 archivePrefix = "arXiv",
 eprint = {1801.01203},
 year = 2018,
 month = jan,
}

@inproceedings{kaiser,
title = "KASLR is Dead: Long Live KASLR",
author = "Daniel Gruss and Moritz Lipp and Michael Schwarz and Richard Fellner and Clémentine Maurice and Stefan Mangard",
year = "2017",
doi = "10.1007/978-3-319-62105-0_11",
isbn = "9783319621043",
volume = "10379 LNCS",
pages = "161--176",
booktitle = "Engineering Secure Software and Systems - 9th International Symposium, ESSoS 2017, Proceedings",
address = "Italy",
}

@article{SgxPectre,
 author = {Guoxing Chen and Sanchuan Chen and Yuan Xiao and Yinqian Zhang and Zhiqiang Lin and Ten H. Lai},
 title = {SgxPectre Attacks: Leaking Enclave Secrets via Speculative Execution},
 journal = {ArXiv e-prints},
 year = 2018,
 month = Feb,
}

@article{meltdownprime,
 author = {Caroline Trippel and Daniel Lustig and Margaret Martonosi},
 title = {MeltdownPrime and SpectrePrime: Automatically-Synthesized Attacks Exploiting Invalidation-Based Coherence Protocols},
 journal = {ArXiv e-prints},
 year = 2018,
 month = Feb,
}

@inproceedings {flush+reload,
author = {Yuval Yarom and Katrina Falkner},
title = {FLUSH+RELOAD: A High Resolution, Low Noise, L3 Cache Side-Channel Attack},
booktitle = {23rd {USENIX} Security Symposium ({USENIX} Security 14)},
year = {2014},
isbn = {978-1-931971-15-7},
address = {San Diego, CA},
pages = {719--732},
url = {https://www.usenix.org/conference/usenixsecurity14/technical-sessions/presentation/yarom},
publisher = {{USENIX} Association},
}

@inproceedings{cache-attacks,
 author = {Osvik, Dag Arne and Shamir, Adi and Tromer, Eran},
 title = {Cache Attacks and Countermeasures: The Case of AES},
 booktitle = {Proceedings of the 2006 The Cryptographers' Track at the RSA Conference on Topics in Cryptology},
 series = {CT-RSA'06},
 year = {2006},
 isbn = {3-540-31033-9, 978-3-540-31033-4},
 location = {San Jose, CA},
 pages = {1--20},
 numpages = {20},
 url = {http://dx.doi.org/10.1007/11605805_1},
 doi = {10.1007/11605805_1},
 acmid = {2117741},
 publisher = {Springer-Verlag},
 address = {Berlin, Heidelberg},
}

@article{cache-missing,
author = {Percival, Colin},
year = {2009},
month = {08},
pages = {},
title = {Cache missing for fun and profit}
}

@article{Tomasulo,
 author = {Tomasulo, R. M.},
 title = {An Efficient Algorithm for Exploiting Multiple Arithmetic Units},
 journal = {IBM J. Res. Dev.},
 issue_date = {January 1967},
 volume = {11},
 number = {1},
 month = jan,
 year = {1967},
 issn = {0018-8646},
 pages = {25--33},
}

@article{NoMo,
 author = {Domnitser, Leonid and Jaleel, Aamer and Loew, Jason and Abu-Ghazaleh, Nael and Ponomarev, Dmitry},
 title = {Non-monopolizable Caches: Low-complexity Mitigation of Cache Side Channel Attacks},
 journal = {ACM Trans. Archit. Code Optim.},
 issue_date = {January 2012},
 volume = {8},
 number = {4},
 month = jan,
 year = {2012},
 issn = {1544-3566},
 pages = {35:1--35:21},
}

@article{CATalyst,
  title={CATalyst: Defeating last-level cache side channel attacks in cloud computing},
  author={Fangfei Liu and Qian Ge and Yuval Yarom and Frank McKeen and Carlos V. Rozas and Gernot Heiser and Ruby B. Lee},
  journal={2016 IEEE International Symposium on High Performance Computer Architecture (HPCA)},
  year={2016},
  pages={406-418}
}

@inproceedings{pcache,
 author = {Wang, Zhenghong and Lee, Ruby B.},
 title = {New Cache Designs for Thwarting Software Cache-based Side Channel Attacks},
 booktitle = {Proceedings of the 34th Annual International Symposium on Computer Architecture},
 series = {ISCA '07},
 year = {2007},
 isbn = {978-1-59593-706-3},
 location = {San Diego, California, USA},
 pages = {494--505},
}

@inproceedings {STEALTHMEM,
author = {Taesoo Kim and Marcus Peinado and Gloria Mainar-Ruiz},
title = {{STEALTHMEM}: System-Level Protection Against Cache-Based Side Channel Attacks in the Cloud},
booktitle = {Presented as part of the 21st {USENIX} Security Symposium ({USENIX} Security 12)},
year = {2012},
isbn = {978-931971-95-9},
address = {Bellevue, WA},
pages = {189--204},
}

@inproceedings{ZhenghongWang,
 author = {Zhenghong Wang and Lee, Ruby B.},
 title = {A Novel Cache Architecture with Enhanced Performance and Security},
 booktitle = {Proceedings of the 41st Annual IEEE/ACM International Symposium on Microarchitecture},
 series = {MICRO 41},
 year = {2008},
 isbn = {978-1-4244-2836-6},
 pages = {83--93},
}

@article{StopWatch,
 author = {Li, Peng and Gao, Debin and Reiter, Michael K.},
 title = {StopWatch: A Cloud Architecture for Timing Channel Mitigation},
 journal = {ACM Trans. Inf. Syst. Secur.},
 issue_date = {November 2014},
 volume = {17},
 number = {2},
 month = nov,
 year = {2014},
 issn = {1094-9224},
 pages = {8:1--8:28},
}

@inproceedings{Liu:2014,
 author = {Liu, Fangfei and Lee, Ruby B.},
 title = {Random Fill Cache Architecture},
 booktitle = {Proceedings of the 47th Annual IEEE/ACM International Symposium on Microarchitecture},
 series = {MICRO-47},
 year = {2014},
 isbn = {978-1-4799-6998-2},
 location = {Cambridge, United Kingdom},
 pages = {203--215},
}

@inproceedings{cloudradar,
  author={Zhang, Tianwei and Zhang, Yinqian and Lee, Ruby B.},
  title={CloudRadar: A Real-Time Side-Channel Attack Detection System in Clouds},
  bookTitle={19th International Symposium on Research in Attacks, Intrusions, and Defenses},
  series = {RAID'16},
  location = {Paris, France},
  year={2016},
}

@misc{horn,
  title     = {{Reading privileged memory with a side-channel}},
  author    = {Jann Horn},
  year      = 2018,
  url  	    = {https://googleprojectzero.blogspot.com/2018/01/reading-privileged-memory-with-side.html}
}

@misc{corbet,
  title     = {{KAISER: hiding the kernel from user space}},
  author    = {Corbet, Jonathan},
  year      = 2018,
  url  	    = {https://lwn.net/Articles/738975/}
}

@inproceedings{multi2sim,
	author = { Ubal, Rafael and Jang, Byunghyun and Mistry, Perhaad and Schaa,
		Dana and Kaeli, David },
	title = {{ Multi2Sim: A Simulation Framework for CPU-GPU Computing }},
	booktitle = { Proc. of the 21st International Conference on Parallel
		Architectures and Compilation Techniques },
	month = { Sep. },
	year = { 2012 }
}

@inproceedings {206170,
	author = {Ferdinand Brasser and Urs M{\"u}ller and Alexandra Dmitrienko and Kari Kostiainen and Srdjan Capkun and Ahmad-Reza Sadeghi},
	title = {Software Grand Exposure: {SGX} Cache Attacks Are Practical},
	booktitle = {11th {USENIX} Workshop on Offensive Technologies ({WOOT} 17)},
	year = {2017},
	address = {Vancouver, BC},
	url = {https://www.usenix.org/conference/woot17/workshop-program/presentation/brasser},
	publisher = {{USENIX} Association},
}

@inproceedings {203183,
author = {Marcus H{\"a}hnel and Weidong Cui and Marcus Peinado},
title = {High-Resolution Side Channels for Untrusted Operating Systems},
booktitle = {2017 {USENIX} Annual Technical Conference ({USENIX} {ATC} 17)},
year = {2017},
isbn = {978-1-931971-38-6},
address = {Santa Clara, CA},
pages = {299--312},
url = {https://www.usenix.org/conference/atc17/technical-sessions/presentation/hahnel},
publisher = {{USENIX} Association},
}

@article{SchwarzWGMM17,
  author    = {Michael Schwarz and
               Samuel Weiser and
               Daniel Gruss and
               Cl{\'{e}}mentine Maurice and
               Stefan Mangard},
  title     = {Malware Guard Extension: Using {SGX} to Conceal Cache Attacks},
  journal   = {CoRR},
  volume    = {abs/1702.08719},
  year      = {2017},
  url       = {http://arxiv.org/abs/1702.08719},
  archivePrefix = {arXiv},
  eprint    = {1702.08719},
  bibsource = {dblp computer science bibliography, https://dblp.org}
}

@inproceedings{Liu:2015:LCS:2867539.2867673,
 author = {Liu, Fangfei and Yarom, Yuval and Ge, Qian and Heiser, Gernot and Lee, Ruby B.},
 title = {Last-Level Cache Side-Channel Attacks Are Practical},
 booktitle = {Proceedings of the 2015 IEEE Symposium on Security and Privacy},
 series = {SP '15},
 year = {2015},
 isbn = {978-1-4673-6949-7},
 pages = {605--622},
 numpages = {18},
 url = {https://doi.org/10.1109/SP.2015.43},
 doi = {10.1109/SP.2015.43},
 acmid = {2867673},
 publisher = {IEEE Computer Society},
 address = {Washington, DC, USA},
}

@inproceedings{weisse2018foreshadowNG,
  title={Foreshadow: Extracting the keys to the intel $\{$SGX$\}$ kingdom with transient $\{$Out-of-Order$\}$ execution},
  author={Van Bulck, Jo and Minkin, Marina and Weisse, Ofir and Genkin, Daniel and Kasikci, Baris and Piessens, Frank and Silberstein, Mark and Wenisch, Thomas F and Yarom, Yuval and Strackx, Raoul},
  booktitle={27th USENIX Security Symposium (USENIX Security 18)},
  pages={991--1008},
  year={2018}
}

@article{islam2019spoiler,
  title={SPOILER: Speculative Load Hazards Boost Rowhammer and Cache Attacks},
  author={Islam, Saad and Moghimi, Ahmad and Bruhns, Ida and Krebbel, Moritz and Gulmezoglu, Berk and Eisenbarth, Thomas and Sunar, Berk},
  journal={arXiv preprint arXiv:1903.00446},
  year={2019}
}

@INPROCEEDINGS{SPLASH, 
author={S. C. Woo and M. Ohara and E. Torrie and J. P. Singh and A. Gupta}, 
booktitle={Proceedings 22nd Annual International Symposium on Computer Architecture}, 
title={The SPLASH-2 programs: characterization and methodological considerations}, 
year={1995}, 
volume={}, 
number={}, 
pages={24-36}, 
doi={10.1109/ISCA.1995.524546}, 
ISSN={1063-6897}, 
month={June},}

@inproceedings{jouppi1998improving,
  title={Improving direct-mapped cache performance by the addition of a small fully-associative cache prefetch buffers},
  author={Jouppi, Norman P},
  booktitle={25 years of the international symposia on Computer architecture (selected papers)},
  pages={388--397},
  year={1998},
  organization={ACM}
}

@inproceedings{HexPADS,
 author = {Payer, Mathias},
 title = {HexPADS: A Platform to Detect "Stealth" Attacks},
 booktitle = {Proceedings of the 8th International Symposium on Engineering Secure Software and Systems - Volume 9639},
 series = {ESSoS 2016},
 year = {2016},
 isbn = {978-3-319-30805-0},
 location = {London, UK},
 pages = {138--154},
 numpages = {17},
 url = {http://dx.doi.org/10.1007/978-3-319-30806-7_9},
 doi = {10.1007/978-3-319-30806-7_9},
 acmid = {2990919},
 publisher = {Springer-Verlag New York, Inc.},
 address = {New York, NY, USA},
}

@article{RSRE,
title = {{ CVE-2018-3640:Rogue System Register Read}},
url = {{https://cve.mitre.org/cgi-bin/cvename.cgi?name=CVE-2018-3640}},
}

@article{SSB,
title = {{ CVE-2018-3639:Speculative Store Bypass}},
url = {https://cve.mitre.org/cgi-bin/cvename.cgi?name=CVE-2018-3639},
}

@article{LFSR,
title = {{ CVE-2018-3665:Lazy FP State Restore}},
url = {https://cve.mitre.org/cgi-bin/cvename.cgi?name=CVE-2018-3665},
}

@article{ret2spec,
  author    = {Giorgi Maisuradze and
               Christian Rossow},
  title     = {ret2spec: Speculative Execution Using Return Stack Buffers},
  journal   = {CoRR},
  volume    = {abs/1807.10364},
  year      = {2018},
  url       = {http://arxiv.org/abs/1807.10364},
  archivePrefix = {arXiv},
  eprint    = {1807.10364},
  bibsource = {dblp computer science bibliography, https://dblp.org}
}

@ARTICLE{smith98, 
author={J. E. {Smith} and A. R. {Pleszkun}}, 
journal={IEEE Transactions on Computers}, 
title={Implementing precise interrupts in pipelined processors}, 
year={1988}, 
volume={37}, 
number={5}, 
pages={562-573}, 
doi={10.1109/12.4607}, 
ISSN={0018-9340}, 
month={May},}

\textbf{Subhash Sethumurugan} is a graduate student at the University of Minnesota – Twin Cities pursuing Phd under Prof. John Sartori. He received a Btech in Electronics and Communication Engineering from NIT Tiruchirappalli. His work concentrates on cache architecture optimizations for performance and security. He also works on application specific optimizations on low power embedded processors.

\textbf{Hari Cherupalli} is a final year PhD Student at the University of Minnesota, working under Prof. John Sartori. He has a B.Tech and an M.Tech in Electrical Engineering from IIT Kharagpur. His research focuses on developing techniques for power management, cost reduction and security in ultra-low-power processors. 

\textbf{Kangjie Lu} is an assistant professor in the Computer Science \& Engineering Department of the University of Minnesota-Twin Cities. He received a Ph.D. in Computer Science from the Georgia Institute of Technology. His current research aims to secure computer systems by hardening code and design, finding vulnerabilities, and detecting privacy leaks.

\textbf{John Sartori} is an associate professor at the University of Minnesota. His research interests include computer architecture, electronic design automation, embedded systems, IoT, wearable technology, and algorithm development, especially focused on energy-efficient computing, ultra-low-power computing, high-performance computing, stochastic computing, and application-aware design and architecture methodologies.

\end{document}